\documentclass[a4paper,11pt]{article}
\pdfoutput=1 

\usepackage{jcappub} 
\usepackage[T1]{fontenc} 
\usepackage{graphicx}

\title{Search for magnetically-induced signatures in the arrival directions of ultra-high-energy cosmic rays measured at the Pierre Auger Observatory}

\author[75]{A.~Aab,}
\author[67]{P.~Abreu,}
\author[50,49]{M.~Aglietta,}
\author[12]{J.M.~Albury,}
\author[1]{I.~Allekotte,}
\author[8,11]{A.~Almela,}
\author[63]{J.~Alvarez Castillo,}
\author[74]{J.~Alvarez-Mu\~niz,}
\author[75]{R.~Alves Batista,}
\author[58,49]{G.A.~Anastasi,}
% ...

\affiliation[1]{Centro At\'omico Bariloche and Instituto Balseiro (CNEA-UNCuyo-CONICET), San Carlos de Bariloche, Argentina}
% ...
\affiliation[8]{Instituto de Tecnolog\'\i{}as en Detecci\'on y Astropart\'\i{}culas (CNEA, CONICET, UNSAM), Buenos Aires, Argentina}
% ...
\affiliation[11]{Universidad Tecnol\'ogica Nacional -- Facultad Regional Buenos Aires, Buenos Aires, Argentina}
\affiliation[12]{University of Adelaide, Adelaide, S.A., Australia}
% ...
\affiliation[49]{INFN, Sezione di Torino, Torino, Italy}
\affiliation[50]{Osservatorio Astrofisico di Torino (INAF), Torino, Italy}
% ...
\affiliation[58]{Universit\`a Torino, Dipartimento di Fisica, Torino, Italy}
% ...
\affiliation[63]{Universidad Nacional Aut\'onoma de M\'exico, M\'exico, D.F., M\'exico}
% ...
\affiliation[67]{Laborat\'orio de Instrumenta\c{c}\~ao e F\'\i{}sica Experimental de Part\'\i{}culas -- LIP and Instituto Superior T\'ecnico -- IST, Universidade de Lisboa -- UL, Lisboa, Portugal}
% ...
\affiliation[74]{Instituto Galego de F\'\i{}sica de Altas Enerx\'\i{}as (IGFAE), Universidade de Santiago de Compostela, Santiago de Compostela, Spain}
\affiliation[75]{IMAPP, Radboud University Nijmegen, Nijmegen, The Netherlands}
% ...
\affiliation[c]{Fermi National Accelerator Laboratory, USA}
% ...

\emailAdd{auger\_spokespersons@fnal.gov}

\abstract{We search for signals of magnetically-induced effects in the arrival directions of ultra-high-energy cosmic rays detected at the Pierre Auger Observatory. We apply two different methods. One is a search for sets of events that show a correlation between their arrival direction and the inverse of their energy, which would be expected if they come from the same point-like source, they have the same electric charge and their deflection is relatively small and coherent. We refer to these sets of events as ``multiplets''. The second method, called ``thrust'', is a principal axis analysis aimed to detect the elongated patterns in a region of interest. We study the sensitivity of both methods using a benchmark simulation and we apply them to data in two different searches. The first search is done assuming as source candidates a list of nearby active galactic nuclei and starburst galaxies. The second is an all-sky blind search. We report the results and we find no statistically significant features. We discuss the compatibility of these results with the indications on the mass composition inferred from data of the Pierre Auger Observatory.}

\begin{document}
\maketitle
\flushbottom

\section{Introduction}
\label{sec:intro}

The identification of the sources of ultra-high-energy cosmic rays (UHECRs) is one of the main unsolved challenges in astrophysics. Since UHECRs are charged particles, their path from their sources to Earth is modified by the extragalactic and Galactic magnetic fields they traverse, most notably by the latter. The knowledge of these intervening fields is still poor, despite the considerable experimental effort in the area (see e.g. \cite{han, beck,durrer} and references therein).

Another element that makes the magnetic deflections difficult to predict is the uncertain composition of UHECRs. This is due to the fact that measurements of the maximum of the shower development, which depends on the mass of the primary particle, have low statistics at the highest energies and its interpretation depends on the modelling of the hadronic interactions. The results from the Pierre Auger Observatory \cite{auger2014a,auger2014b,augerxmax2019} indicate that the composition becomes heavier with increasing energy. However, measurements do not rule out a light nuclei fraction at the highest energies, which might originate in a few nearby sources, different from the average ones. In such a case, the search for magnetically-induced signatures in the arrival directions of UHECRs could help identify this type of sources. Several analysis techniques have been designed to capture this kind of effect, including indirect ways such as in~\cite{JPL-UHECR2018}. In this work, we show the results of two different methods applied to perform this search in a direct way, which are referred to as ``multiplets'' and ``thrust''.

Multiplets are defined as a set of events that show a correlation between their arrival direction and the inverse of their energy, which is expected if they come from the same source, they have the same electric charge and their deflections are small and remain coherent. The observation of multiplets could enable the accurate identification of the direction of the source and could also provide a new means to probe the Galactic magnetic field by inferring the value of its integrated component orthogonal to the trajectory of cosmic rays.

A search for multiplets was performed by the Pierre Auger Collaboration in \cite{auger2012}. In that work, data up to 31st December 2010 were used, which amounted to a total exposure of 25,800\,km$^2$~sr~yr. The results obtained were not statistically significant. The largest multiplet found above 20\,EeV had 12 events and the probability that it would appear by chance in an isotropic distribution of events was $6\%$. There were also two independent 10-plets and the chance probability of having at least three multiplets with a multiplicity equal to or larger than 10 was $20\%$. With the larger dataset used in this work, the number of events added to these multiplets is not statistically significant when comparing it to the number of events that would be added if the arrival directions were isotropic, with a $p$-value of $60.5\%$.

In the thrust method, an observable is built from a principal axis analysis in a localized region of the sky, measuring the elongation of a pattern with respect to the center of the region of interest (ROI). The thrust method was applied to data by the Pierre Auger Collaboration in \cite{auger2015}, with events detected up to 19th March 2013, amounting to a total exposure of 32,800\,km$^2$~sr~yr. The measured distributions of the thrust observables with the centroid corresponding to the highest energy cosmic rays did not reveal any local patterns in the arrival directions of UHECRs.

In this work, we update these analyses with more statistics. The data set used, which amounts to a total exposure of 101,900\,km$^2$~sr~yr, is described in Section~\ref{sec:data}. This is an increase of a factor 4 with respect to \cite{auger2012} and a factor 3 with respect to \cite{auger2015}. The methods used are discussed in detail in Section~\ref{sec:methods}. Their expected sensitivity is shown in Section~\ref{sec:sensi}, using a benchmark simulation described in Section~\ref{sec:simu}. The methods are applied to a targeted search on a selection of nearby active galactic nuclei (AGNs) and starburst galaxies (SBGs), candidates to be sites of ultra-high-energy (UHE) acceleration, and the results are presented in Section~\ref{sec:targeted}. Moreover, in Section~\ref{sec:blind}, we apply the methods to an all-sky blind search. Finally, in Section~\ref{sec:concl}, we present our conclusions.

\section{The Pierre Auger Observatory and the data set}
\label{sec:data}
The Pierre Auger Observatory \cite{augernim04,augernim15}, located in Malarg\"ue, Argentina  (35.2$^\circ$~S, 69.5$^\circ$~W, 1400\,m~a.s.l.), consists of a large surface detector (SD) and an air-fluorescence detector (FD), which are used to observe extensive air showers generated by UHECRs in a complementary way. The SD is composed of an array of 1660 water-Cherenkov stations, arranged in an equilateral triangular grid, spread over an area of 3000\,km$^2$. It measures, with a duty cycle of nearly 100\%, the energy flow in the shower carried by electrons, photons and muons that reach ground level. The FD, comprising 27 telescopes at four sites, overlooks the surface array and observes, with a duty cycle close to 15\%, the fluorescence light emitted by nitrogen molecules excited by the particles produced as the air shower develops in the atmosphere.

In this study, we select data recorded with the surface detector between 1st January 2004 and 31st August 2018 with zenith angles $\theta\leq 80^\circ$. The energy thresholds we use are 20\,EeV and 40\,EeV, which yield data sets of 6568 and 1119 events, respectively. The ``vertical'' events ($\theta\leq 60^\circ$) are required to have at least four active stations surrounding the one with the highest signal, while the ``inclined'' events ($\theta > 60^\circ$) are required to have at least five active stations. For the vertical events, the reconstructed core must be inside an equilateral or isosceles triangle of active stations. This ensures an accurate reconstruction of the shower geometry and energy.

The arrival directions of these cosmic rays are determined from the relative arrival times of the shower front in the triggered stations. The angular resolution, defined as the radius around the true cosmic-ray direction that would contain 68\% of the reconstructed shower directions, is better than 1$^\circ$ for the energies considered here \cite{augerar,augerar2019}. The energy estimate is obtained from the signals recorded by the SD stations \cite{augerrec1,augerrec2} and is calibrated using ``hybrid'' events (detected simultaneously by SD and FD) and taking advantage of the quasi-calorimetric energy determination obtained with the FD \cite{augerE2019}. The statistical uncertainty in the energy determination is smaller than 9\% for the energies considered and the systematic uncertainty on the absolute energy scale is 14\% \cite{augerE2019}.

At energies above 4\,EeV, the SD trigger is fully efficient, and the determination of the exposure is purely geometrical \citep{augernim10}. For the time period considered and for the applied energy and zenith angle selection, it amounts to 101,900\,km$^2$~sr~yr.

\section{Methods}
\label{sec:methods}
\subsection{Multiplets}
If the deflection is small, it is a good approximation \cite{multiGHRM} to consider that the relation between the arrival direction of the cosmic ray $\vec{\Theta}$ and the actual source direction $\vec{\Theta}_\text{s}$,
\begin{equation}
 \vec{\Theta} = \vec{\Theta}_\text{s} +  \frac{Ze}{E} \int^{L}_{0} {\rm d} \vec{\ell} \times \vec{B}(\vec{\ell})  \,.
\end{equation}
This equation can be simplified to yield the following linear relation with the inverse of the energy $E$ of the cosmic ray:
\begin{equation}
 \vec{\Theta} \simeq  \vec{\Theta}_\text{s} + \frac{\vec{D}(\vec{\Theta}_\text{s})}{E}  \, .
\end{equation}
Here $\vec{D}$ is the integral along the line of sight of the perpendicular component of the magnetic field $\vec B$ times the charge $Ze$ of the particle, and is approximated to a constant for a fixed source direction. The ``deflection power'' $D \equiv|\vec{D} |$ will be given in units $1^{\circ} \times 100$\,EeV.

To determine if a set of cosmic rays forms a multiplet, we first use a coordinate system ($x,y$) in the tangent plane to the celestial sphere (centered on the average direction of the events). The system is then rotated to a system ($u,w$) with an angle such that the covariance
\begin{equation}
 \text{Cov}(w, 1/E) = \frac{1}{N} \sum_{i=1}^N (w_i - \langle w \rangle) (1/E_i -\langle 1/ E \rangle)
\end{equation}
between the cosmic ray coordinates $w_i$ and their inverse energy $1 / E_i$ is zero and the covariance $\text{Cov} (u, 1/E)$ is maximal. This angle is given by
\begin{equation}
 \alpha = \arctan \left(\frac{\text{Cov}(y,1/E)}{\text{Cov}(x,1/E)}\right).
\end{equation}
The correlation between the coordinate $u$ and $1/E$ is measured through the correlation coefficient
\begin{equation}
 C(u, 1/E) = \frac{\text{Cov}(u, 1/E)}{\sqrt{\text{Var}(u) \text{Var}(1/E)}}
\end{equation}
where the variances are given by $\text{Var}(x) = \left\langle (x - \langle x \rangle)^2 \right\rangle$.

The cuts applied to identify possible candidates among background chance alignments are: i) a minimum value for the correlation coefficient, $ C(u, 1/E) > C_{\text{min}}$, and ii) a maximum value for the spread in the transverse direction of the multiplet, $\max(|w_i - \langle w\rangle|) < W_{\text{max}}$. In \cite{auger2012}, the two parameters chosen were $C_{\text{min}} = 0.9$ and $W_{\text{max}} = 1.5^{\circ}$. They were adopted following a search for a compromise between maximizing the signal from a true source and minimizing the background arising from chance alignments. To do so, numerical simulations were performed with sets of protons from randomly-located extragalactic sources, which were propagated through a bisymmetric magnetic field with even symmetry and a simple random deflection with root mean square amplitude $1.5^\circ\times$(20\,EeV/$E$) to simulate the turbulent field. In this work, we tested variations of these parameters using the more up-to-date simulation described in Section~\ref{sec:simu} and, as it is shown in Section~\ref{sec:sensi-mplets}, these values for the cuts still perform well. A third cut used here consists in searching for multiplets which extend up to $20^\circ$ in the sky and with energies above 40\,EeV, so that the linear correlation of the deflection with $1/E$ is expected to be valid for ``light'' nuclei, such as emitted by proton and helium sources. We also test the sensitivity of the method making a search of events with energies between 20\,EeV and 40\,EeV, where the linear approximation is still valid for proton sources. 

\subsection{Thrust ratio}
\label{sec:thrust}

Deviations from a clear correlation between the energy and the deflection angle can arise from large turbulent field components or from a mixed nuclear composition of the cosmic rays, as deflections scale also with the nuclear charges $Z_i$ of the particles. In this case, the thrust ratio, $T_2 / T_3$, can still capture an elongation of the arising pattern in the arrival directions. The thrust observables $T_k$ $(k=1,2,3)$ are constructed by successively maximizing the values
\begin{equation}
T_k = \underset{\hat{n}_k}{\text{max}} \left(
 \frac{\sum_i | \omega_i^{-1} \, \vec p_i \cdot \hat{n}_k |}
{\sum_i | \omega_i^{-1} \, \vec p_i | } \right)
\end{equation}
with respect to the axes $\hat{n}_k$ starting with $k=1$. The sum iterates over all cosmic rays in the chosen region of interest (ROI), $\vec p_i$ is the arrival direction of particle $i$ weighted with its energy and $\omega_i$ is the exposure towards this direction. The principal axes $\hat{n}_k$ are perpendicular to each other ($\hat{n}_1 \perp \hat{n}_2 \perp \hat{n}_3$), thus, we obtain by construction for the thrust observables the relation $T_1 > T_2 > T_3$.

While the thrust axis $\hat{n}_1$ points radially to the local barycenter of the energy distribution within the ROI, the axes $\hat{n}_2$ and $\hat{n}_3$ form an orthonormal coordinate system tangential to the local spherical plane. In this system, a signal-like structure would be characterized by a strong collimation of arrival directions around the $\hat{n}_2$-axis, and therefore by a high thrust observable $T_2$. To be less affected by the radial cosmic-ray distribution within the ROI, we choose instead the thrust ratio $T_2 / T_3$ as the relevant observable. Based on the simulations performed in Section \ref{sec:simu}, we choose for the ROI radius a value of $r_{\text{ROI}} = 0.3$~rad~(${\sim} 17.2^\circ$). A more detailed description of the procedure can be found in \cite{auger2015}.

\section{Target selection and benchmark simulation}
\label{sec:simu}
In this Section, we describe the selection of astrophysical objects that have been discussed as promising candidates for UHE acceleration and which are used in a targeted search. We also give details of the simulation we used to evaluate the sensitivities of the two methods presented in the previous Section.
\subsection{Target selection} \label{sec:sources}
In \cite{auger2018}, we reported on a potentially interesting indication of anisotropy in the arrival direction of UHECRs given by an excess flux towards nearby galaxies. A maximum-likelihood ratio test was used with two free parameters, namely the fraction of events correlating with astrophysical objects and an angular scale characterizing the clustering of cosmic rays around extragalactic sources. Sky models of cosmic-ray density were constructed using two catalogs of source classes studied by \emph{Fermi}-LAT \cite{2FHL,fermi2012}: AGNs and SBGs. The relative luminosity of each source in UHECRs was assumed to be proportional to the $\gamma$-ray luminosity for the former and to the radio emission for the latter, for which the sample of detected sources in the $\gamma$-ray band remains scarce. When comparing both models to isotropic expectations, the most promising result was found with SBGs, with a statistical significance of $4.0\sigma$. The result found with AGNs was also interesting, with a $2.7\sigma$ deviation from isotropy. Results were updated in \cite{updateSBGICRC} and the statistical significances increased to $4.5\sigma$ and $3.1\sigma$ for the SBG and AGN models, respectively. The Telescope Array Collaboration tested these results with their observed events, but they found not to have enough statistics to confirm these results \cite{TASBG}.

In this work, we include a search for magnetically-induced effects in the arrival directions of events close to potential candidates for UHE acceleration. For the multiplet search, we look for sets of events with a reconstructed source position within $3^\circ$ of these candidates. We also apply the thrust method centering the ROI in the position of the source candidates. We call these two analyses a ``targeted search''.

\begin{figure}
\centering \includegraphics[width=0.6\textwidth]{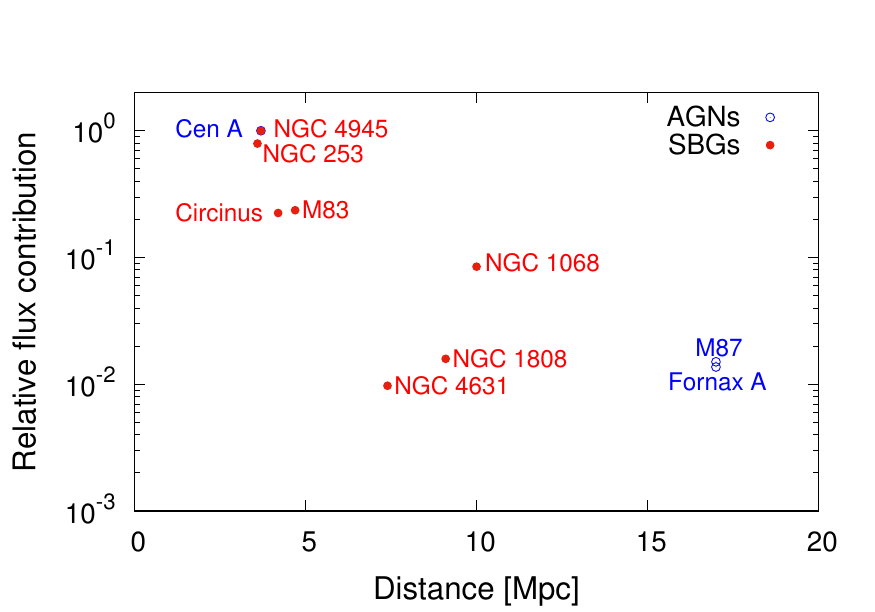}
\caption{Relative flux with respect to the brightest object in each catalog (AGNs in blue empty circles and SBGs in red filled circles) as a function of the distance to the source candidate, accounting for the directional exposure of the Observatory and assuming that the accelerated and detected events are helium nuclei with an energy greater than 40\,EeV. The results are marginalized over the uncertainties on the measured energy and on the distance to the candidate.}
\label{fig:source_cand}
\end{figure}

To define the astrophysical objects that are chosen as source candidates, we used the catalogs from \cite{2FHL,fermi2012,fermi2017,becker}. We divided the sample between AGNs and SBGs. Since we are interested in ``light'' UHECRs, we computed the flux at Earth as if these candidates would accelerate helium nuclei, accounting for the interactions from the sources to Earth. If with the methods used in this work we were able to detect such events, we would also be able to detect protons in the case they were also accelerated at these energies. In Figure \ref{fig:source_cand}, we show the flux contribution of these sources relative to the expected one from the brightest source in each sample, as a function of the distance to the object, taking also into account the non-uniform exposure on the celestial sphere of the Observatory. We simulated events with an energy above 40\,EeV and with an energy spectrum at the source being ${\rm d}N/{\rm d}E \propto E^{-2}$. We assume that these sources, different from the average ones, would contribute to a small fraction of the total measured flux. Hence, the spectrum assumed for these simulations would be compatible to the measured one. We present the results marginalized over the uncertainties on the measured energy and on the distance to the candidate.\footnote{The distance estimates employed in this work are based on \cite{tully2013} and \cite{tully2008}, exploiting in particular the Tully-Fisher correlation or the tip of the red-giant branch. The typical uncertainty on high-quality distance estimators is estimated to 10\% \cite{2009arXiv0902.3668T}, which we adopt as the minimum uncertainty in the marginalization procedure. The largest variance in distance estimates over different methods is found for NGC\,1068, for which redshift-based estimates go up to 19\,Mpc \cite{Cappellari}. Adopting the latter distance for NGC\,1068 does not impact the selection of this source within the sample of candidate for emitters of elongated patterns of UHECR.} We selected the candidates using a cut on the helium flux contribution of $1 \%$ relative to the object with the highest contribution (Cen\,A and NGC\,4945 for the AGN and SBG catalogs, respectively). We also kept NGC\,4631 since its contribution, $0.96\%$, is barely below the threshold. In Table \ref{tab:targets}, we list the ten source candidates that remain after this cut, which include three AGNs and seven SBGs.

\hspace{1cm}
\begin{table}[h]
    \centering
    \caption{Selection of astrophysical objects which could be candidates for UHE acceleration. Assuming that these candidates would accelerate helium nuclei, these objects are expected to contribute at least ${\sim} 1 \%$ with respect to the strongest source candidate in each catalog, accounting for propagation effects and the exposure of the Observatory.}
    \label{tab:targets}
    \vspace{0.25cm}
    \begin{tabular}{ |p{3cm}||p{3cm}|p{3cm}|p{3cm}|  }
    \hline
    \multicolumn{4}{|c|}{AGN source candidates}\\
    \hline
    Target & Gal. longitude [$^{\circ}$] & Gal. latitude [$^{\circ}$] & Distance [Mpc] \\
    \hline
    Cen\,A &  309.5 & $\phantom{+}19.4$ & $\phantom{0}3.7$ \cite{tully2013}\\
    M\,87 &  283.8  & $\phantom{+}74.5$ & $17$ \cite{tully2013}\\
    Fornax A & 240.2 & $-56.7$ &  $17$ \cite{tully2013}\\
    \hline
    \hline
    \multicolumn{4}{|c|}{SBG source candidates}\\
    \hline
    Target & Gal. longitude [$^{\circ}$] & Gal. latitude [$^{\circ}$] & Distance [Mpc] \\
    \hline
    NGC\,253  &  $\phantom{0}97.4$  &  $-88.0$  &  $\phantom{0}3.6$ \cite{tully2013} \\
    NGC\,4945  &  305.3  &  $\phantom{+}13.3$  & $\phantom{0}3.7$ \cite{tully2013}  \\
    Circinus  &  311.3  &  $-3.8$  &  $\phantom{0}4.2$  \cite{tully2008} \\
    M\,83  &  314.6  &  $\phantom{+}32.0$  &  $\phantom{0}4.7$ \cite{tully2013} \\
    NGC\,4631  &  142.8  &  $\phantom{+}84.2$  &  $\phantom{0}7.4$  \cite{tully2013}\\
    NGC\,1808  &  241.2   & $-35.9$  &  $\phantom{0}9.1$ \cite{tully2013} \\
    NGC\,1068   & 172.1   & $-51.9$  &  $10$  \cite{tully2008}\\
    \hline
    \end{tabular}
\end{table}

\subsection{Benchmark simulation}
To evaluate the performance of the methods presented in Section~\ref{sec:methods}, we tested their sensitivity in a model where the UHECRs are accelerated in the source candidates presented in Section~\ref{sec:sources} (listed in Table \ref{tab:targets}) and then are propagated through a model of the intervening magnetic fields. For the propagation through the Galactic magnetic field, we use a lensing technique where particles of negative charge are backtracked from Earth to outside the Galaxy \cite{bretz2014}.

Extragalactic and Galactic magnetic fields deflect the UHECRs from their sources to Earth. The former fields are the least known, but assuming that they have a strength smaller than ${\sim} 10^{-9}$\,G and a correlation length smaller than $1$\,Mpc \cite{Kronberg:1993vk,durrer,Pshirkov:2015tua,Subramanian}, the deflections of helium nuclei coming from the considered candidates at a distance less than 20\,Mpc are negligible with respect to the deflections that occur during the propagation within our Galaxy.

The Galactic magnetic field can be further divided into regular and turbulent components. The regular part of the Galactic magnetic field is expected to give a dominant contribution to the UHECR deflections. There are several models of the regular Galactic field in the literature and there are large uncertainties (see e.g. \cite{farrar,erdmann,farrarunger}) so that their predictions cannot yet be used directly as an ingredient in the analysis. To test the sensitivities of the methods in this work, we chose to use the regular model of Jansson and Farrar \cite{jansson2012a}, and two different scenarios for the turbulent component. One, from \cite{jansson2012b}, has a ``striated'' random field (having the same orientation as the local coherent field but with a strength and sign which varies on a small scale) and a random Kolmogorov field with coherence length of $60$~pc (referred to as ``GMF-A'' model). The model is empirical and data-driven, and the striated poloidal, toroidal and spiral components allow for effects such as a Galactic wind and the compression of an isotropic turbulent field. A second model for the random field is based on the analysis of the Planck team, which argues that the synchrotron emission map of WMAP used by \cite{jansson2012b} to fix the strength of the turbulent component of GMF-A is likely too large \cite{planck2016}, and hence rescales the random field of \cite{jansson2012b} correspondingly. Therefore, we also simulate a weaker version of the turbulent field with the same coherence length but strength scaled-down by a factor $3$, and without the striated field; we call this the ``GMF-B'' model. 

To evaluate the sensitivity of the methods, we test scenarios where one of the source candidates, listed in Table \ref{tab:targets}, accelerates a certain number of cosmic rays, $N_\text{s}$, with an energy spectrum following a simple power law with $E^{-2}$. Two different scenarios are assumed for the chemical composition of the accelerated particles. In Scenario 1, the source accelerates helium nuclei with an energy between 40\,EeV and 200\,EeV. In Scenario 2, we assume that helium nuclei are accelerated with energies between 40\,EeV to 80\,EeV, and that there is a proton fraction between 20\,EeV to 40\,EeV. The number of accelerated helium nuclei is chosen to be, arbitrarily in this example, twice as high as the number of the proton events. These energy ranges were chosen assuming that the acceleration at the source follows a simple rigidity-dependent mechanism as expected from acceleration by electromagnetic fields. In Table \ref{tab:scenarios}, we show a brief summary of the chosen scenarios.

In both scenarios, we propagate the cosmic rays from the sources through GMF-A and GMF-B models. Hence, there are four simulation setups to be analyzed. We simulate a total number of events\footnote{These numbers are slightly smaller than the actual number of events in data, but the difference is expected to have a minor impact on the sensitivity results that are presented.} $N_{\rm tot}=900$ above 40\,EeV in Scenario 1, and $N_{\rm tot}=6000$ above 20\,EeV in Scenario 2. In each simulation, each source candidate is studied separately, there are $N_s$ events originating in the chosen object, plus $N_{\rm tot}-N_\text{s}$ isotropically distributed events, which follow the geometrical exposure of the Observatory and the measured energy spectrum. The arrival directions and energies of the $N_\text{s}$ events are simulated accounting for the experimental resolutions. Thus, we apply an experimental energy uncertainty\footnote{This is a worst-case scenario compared to the energy uncertainty reported in \cite{augerE2019}, which was found to be smaller than 9\% for the energies considered, by making improvements in the calibration, atmospheric treatment, reconstruction, and invisible energy.} of $14 \%$ and an angular resolution of $1^{\circ}$.

\begin{table}[h]
    \centering
    \caption{Summary of the details of the scenarios chosen to study the sensitivity of the methods.}
    \label{tab:scenarios}
    \vspace{0.25cm}
   \begin{tabular}{ | p{3cm}||p{3cm}|p{4.5cm}|  }
    \hline
    \ & \multicolumn{1}{|c|}{Scenario 1}  & \multicolumn{1}{|c|}{Scenario 2} \\
    \hline
    Minimum Energy   &  \multicolumn{1}{|c|}{40\,EeV} & \multicolumn{1}{|c|}{20\,EeV} \\
    Maximum Energy   &  \multicolumn{1}{|c|}{200\,EeV}  & \multicolumn{1}{|c|}{80\,EeV} \\
    Composition      & \multicolumn{1}{|c|}{He} & p~for~ $E<40\,\text{EeV}$, He~for~$E>40\,\text{EeV}$\\
    GMF models       & \multicolumn{1}{|c|}{A and B} & \multicolumn{1}{|c|}{A and B} \\
    \hline
    \end{tabular}
\end{table}

\begin{figure}
\includegraphics[width=0.5\textwidth]{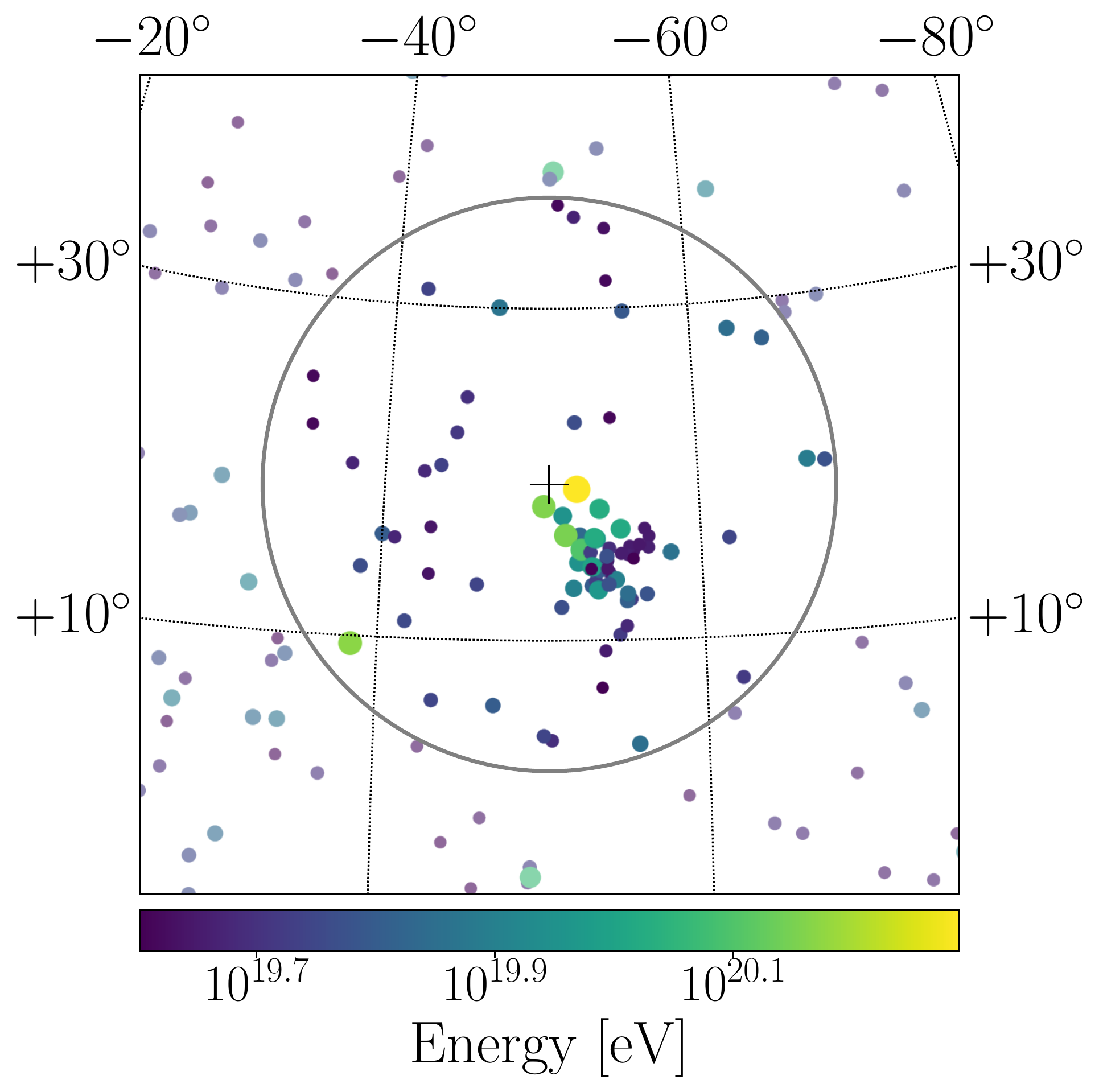}
\includegraphics[width=0.5\textwidth]{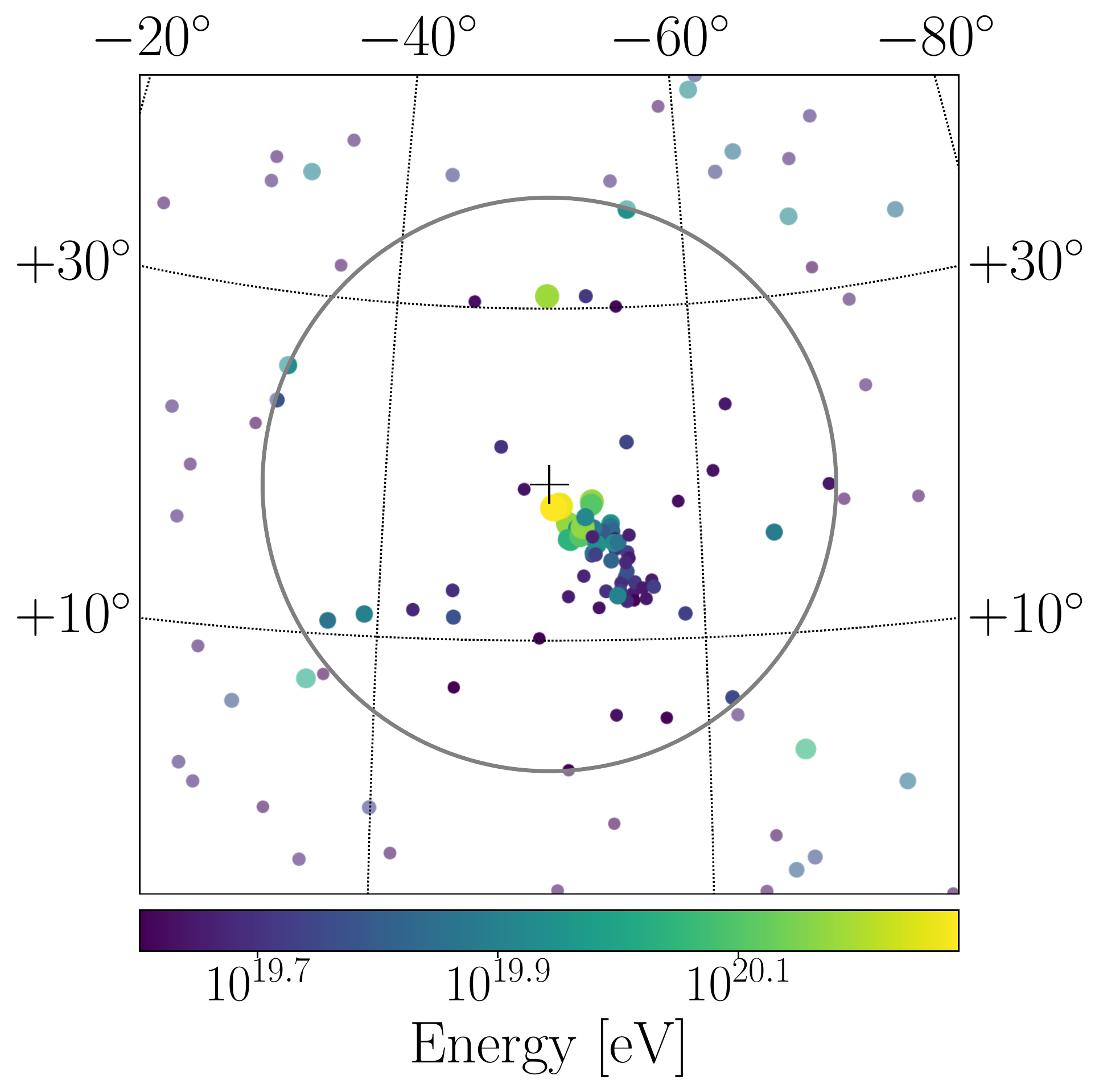}
\caption{Examples of two simulations with $N_\text{s}=50$ events from Cen\,A in Scenario 1, for the GMF-A (left) and the weaker version GMF-B (right), in Galactic coordinates.}
\label{fig:benchmark_simulation}
\end{figure}

In Figure \ref{fig:benchmark_simulation}, we show two examples of a simulation in Scenario 1, with Cen\,A as the source candidate, a number of injected signal cosmic rays of $N_\text{s}=50$ and both Galactic magnetic field models GMF-A (left panel) and GMF-B (right panel), in Galactic coordinates.

\section{Sensitivity studies}
\label{sec:sensi}

\subsection{Sensitivity of the multiplets method}
\label{sec:sensi-mplets}

In this Section, we show the results of the sensitivity studies with the multiplet method, assuming the acceleration Scenario 1 (described in the previous Section \ref{sec:simu}), where helium nuclei are accelerated with energies between 40\,EeV and 200\,EeV. We perform $1000$ simulations where a given $N_\text{s}$ source events are simulated from each source candidate.

For each simulation of $N_\text{s}$ events, the significance of the reconstructed multiplet is calculated by computing the fraction of  simulations of isotropically distributed events (following the geometric exposure of the Observatory), with the same total number of events $N_{\rm tot}$ and with the same energy spectrum, in which a multiplet with the same or larger multiplicity and passing the same cuts appears by chance.

When injecting $N_\text{s}$ events, the reconstructed multiplet does not necessarily have a multiplicity equal to $N_\text{s}$, since there can be simulated events that do not pass the required cuts, and there can be isotropically distributed events which align by chance with simulated ones from the chosen source. To account for this fact, in the sensitivity studies we present the mean chance probabilities computed as:
\begin{equation}
\left < P(N_\text{s}) \right > = \sum_n f(n)P_{\rm ch}(n),
\end{equation}
where $f(n)$ is the fraction of $n$-plets obtained after having simulated $N_\text{s}$ source events and $P_{\rm ch}(n)$ is the isotropic chance probability for a multiplet with $n$ or more events.

\begin{figure}
\includegraphics[width=0.48\textwidth]{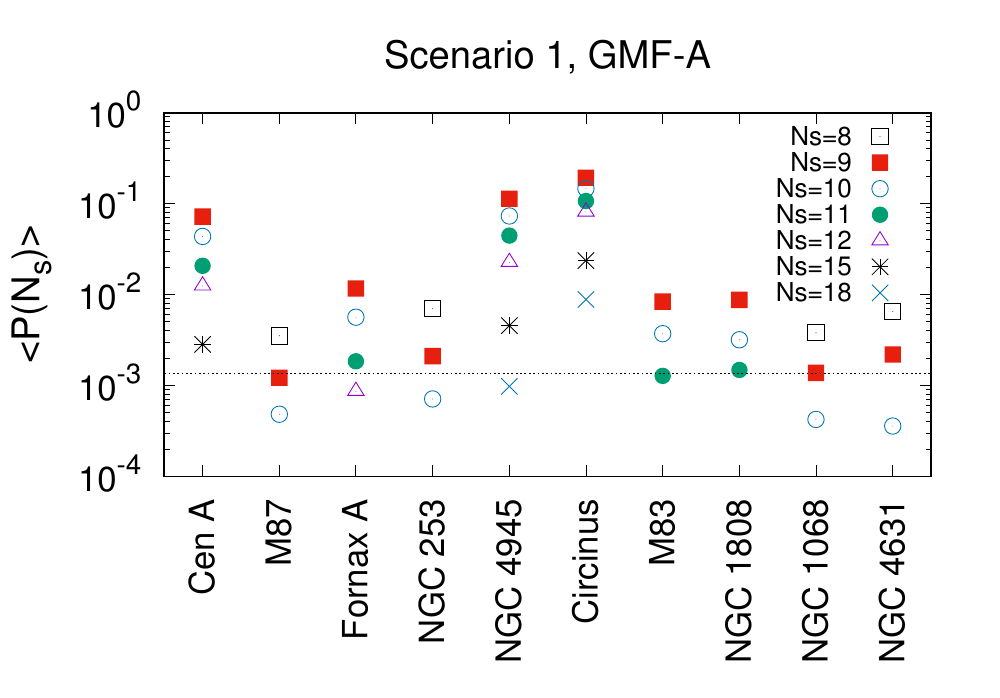}
\includegraphics[width=0.48\textwidth]{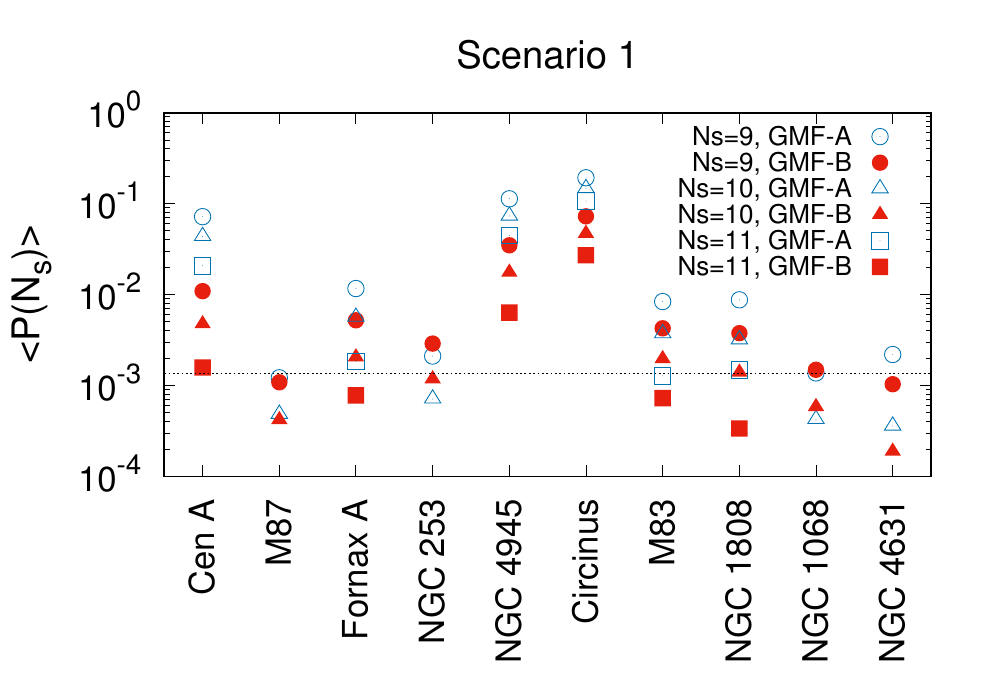}
\caption{Sensitivity results for the multiplet method in Scenario 1 ($E>40$~EeV). Mean chance probability $\left < P(N_\text{s}) \right >$ for different number of injected cosmic rays $N_\text{s}$ as a function of the chosen source, for GMF-A (left panel) and a comparison between GMF-A and GMF-B (right panel). The dashed line indicates the one-sided $3\sigma$ chance probability.}
\label{fig:sensitivity_multiplet}
\end{figure}

In the left panel of Figure \ref{fig:sensitivity_multiplet}, we present the mean chance probabilities for different cases of $N_\text{s}$ source events (shown with different markers), for the ten source candidates considered, in Scenario 1 and with the source events propagated through the GMF-A model. Due to the non-uniform exposure of the Observatory and to the dependence of the deflections and of the magnification in the Galactic magnetic field on the location of the source \cite{hmr,FarrarSutherland}, the number of events from the source required to obtained a mean chance probability  corresponding to a $3 \sigma$ level depends highly on the source candidate. For example, $N_\text{s}=9$ injected cosmic rays for the sources M\,87 and NGC\,1068 are needed to reach a $3 \sigma$ level, while $N_\text{s}=18$ injected cosmic rays are needed for the source NGC\,4945. This corresponds to signal fractions of $1 \%$ and $2 \%$, respectively.

In the right panel of Figure \ref{fig:sensitivity_multiplet}, we show a comparison of the two Galactic magnetic field models, GMF-A and GMF-B, for $N_\text{s}$ from 9 to 11. As expected, the multiplet method performs better for almost all source candidates with the GMF-B model, where the random deflections are weaker.

The values obtained for the number of events from the sources are reasonable. For instance, if we use the results published in \cite{auger2018} for the AGN and SBG models, the anisotropic fractions are $(7 \pm 4)\%$ and $(10 \pm 4)\%$, respectively. In the former case, if the three AGNs listed in Table \ref{tab:targets} are the largest contributors, and if we assume that their relative flux contribution is the one shown in Figure \ref{fig:source_cand}, their respective contribution would be $6.8\%$ for Cen\,A and ${\sim}0.1\%$ for M\,87 and Fornax\,A. Translating these fractions to number of source events, it would mean $N_\text{s} \approx 61$ for Cen\,A and $N_\text{s} \approx 0.8$ and $N_\text{s}\approx 0.9$ for M\,87 and Fornax\,A, respectively. If we compare to the sensitivity studies done for instance with GMF-A model (see the left panel of Figure \ref{fig:sensitivity_multiplet}), to obtain a mean $p$-value corresponding to a $3\sigma$ level, one would need $N_\text{s} \approx 16$ for Cen\,A, $N_\text{s} \approx 9$ for M\,87 and $N_\text{s} \approx 12$ for Fornax\,A. Though the results are not as favorable for M\,87 and Fornax\,A as they are for Cen\,A, their relative luminosity in UHECRs might not be proportional to their $\gamma$-ray luminosity as we assume here \cite{2018MNRAS.479L..76M}.

A similar comparison can be done with the SBG model. The number of events from the sources needed for a $3\sigma$ detection are well within the expectations from the results of \cite{auger2018} for the largest contributors, NGC\,4945 and NGC\,253 and not as much for NGC\,1808 and NGC\,4631, where less than one event is expected and $N_\text{s}$ between 9 and 11 are needed (as it happens in the case of M\,87 and Fornax\,A).

We should point out that the angular scales of $7^\circ$ and $13^\circ$, where the maximum deviation from isotropy was found in \cite{auger2018} for the AGN and SBG models, could be compatible with a He composition given the uncertainties on the GMF model. For example, using the simulations from Scenario 1, for all sources, at least $95\%$ of the events would be within the angular window for the SBG model. In the case of the AGN model, at least $40\%$ of the events would be contained within such an angular window for the ten sources, with six sources having a containment of at least $85\%$. 

\begin{figure}
\centering \includegraphics[width=0.48\textwidth]{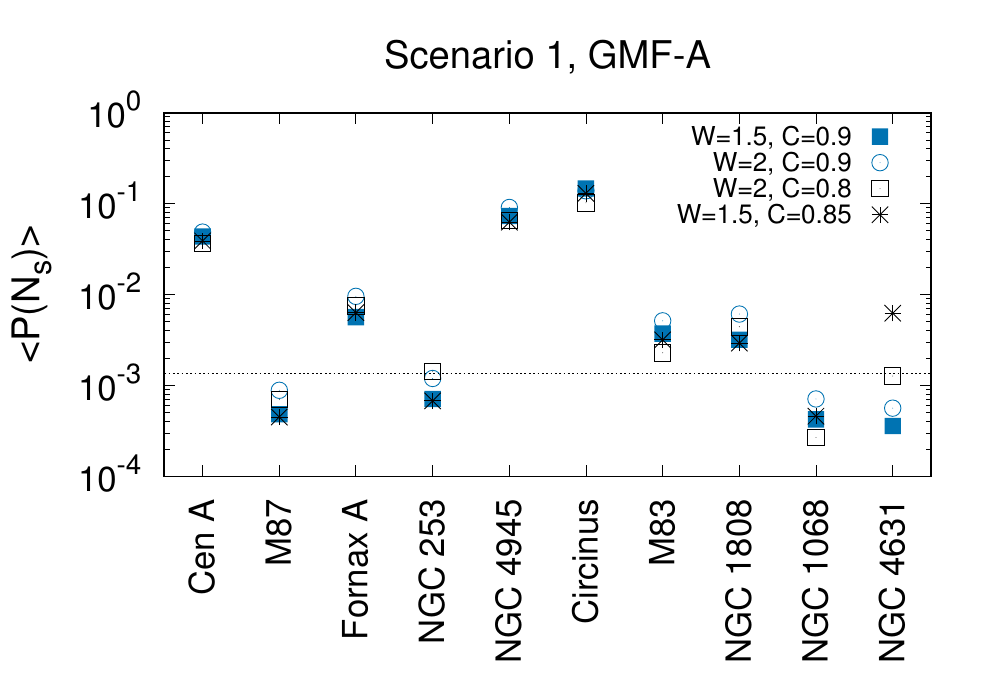}
\caption{Mean chance probability $\left < P(N_\text{s}) \right >$ for $N_\text{s}=10$ as a function of the chosen source, for the GMF-A model and in Scenario~1. The different markers show the results for different values of the cuts  $C_{\text{min}}$ and $W_{\text{max}}$. The dashed line indicates the one-sided $3\sigma$ chance probability.}
\label{fig:sensitivity_multiplet_cuts}
\end{figure}

Using the simulation of Scenario 1 with the GMF-A model and a number of injected source events of $N_\text{s}=10$, we tested variations of the cuts on the correlation coefficient $C_{\text{min}}$ and on the spread in the transverse direction of the multiplet $W_{\text{max}}$. The mean chance probability $\left < P(N_\text{s}) \right >$ as a function of the chosen source, is shown in Figure \ref{fig:sensitivity_multiplet_cuts}. The original cuts of $C_{\text{min}}=0.9$ and $W_{\text{max}}=1.5^\circ$ perform best for most sources. When making the cuts looser, more of the injected signal events can be recovered in the reconstructed multiplet. However, the significance of a larger multiplet passing looser bounds is not necessarily better than the significance of the smaller multiplet with stricter cuts, since in the former case, the probability of chance alignments in the isotropically distributed events also increases.

\begin{figure}
\centering \includegraphics[width=0.48\textwidth]{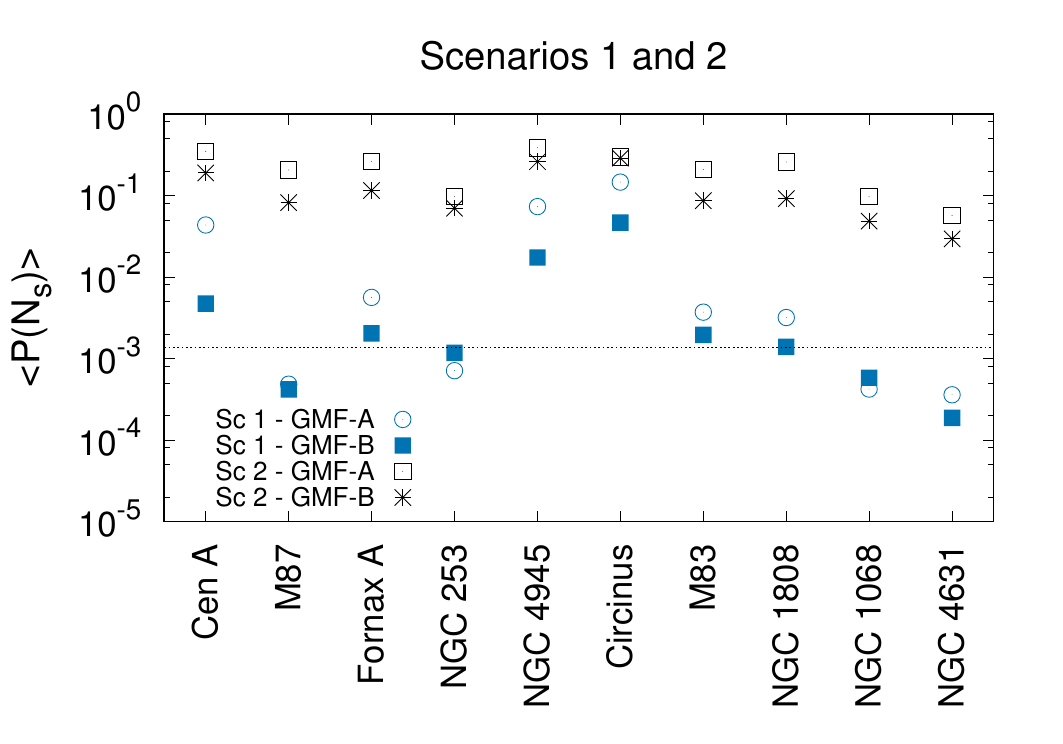}
\caption{Mean chance probability $\left < P(N_\text{s}) \right >$ as a function of the chosen source. The black empty squares and asterisks are the results for Scenario 2, with $N_\text{s}=10$ above 40\,EeV and $N_s=5$ between 20\,EeV and 40\,EeV, for the two GMF models considered, respectively. The blue filled squares and empty circles are the results for Scenario 1, with $N_\text{s}=10$ above 40\,EeV, for the two GMF models considered, respectively. The dashed line indicates the one-sided $3\sigma$ chance probability. }
\label{fig:sensitivity_multiplet_scenario2}
\end{figure}

We studied the sensitivity of the method with Scenario 2. To perform the search in this scenario, we first searched for multiplets with events with energies above 40\,EeV. After that, we selected the largest multiplet and we searched for events between 20\,EeV and 40\,EeV which would correlate with the events above 40\,EeV. To pre-select these events, we chose those at an angular distance to the event with the highest energy in the multiplet expected if the deflection power was the one obtained with the higher energy events, with an uncertainty of $20\%$, assuming that the events between 20\,EeV and 40\,EeV were protons and those above 40\,EeV were He. We once more applied a cut to the maximum angular distance of $20^\circ$ so that the linear approximation between $1/E$ and the arrival directions is still expected to be valid. We simulated $N_\text{s}=10$ source events from each source candidate with energies above 40\,EeV and $N_\text{s}=5$ source events with energies between 20\,EeV and 40\,EeV and we compared the results to those obtained with Scenario 1 with the same number of source events $N_\text{s}=10$ above 40\,EeV. We present the results in Figure \ref{fig:sensitivity_multiplet_scenario2}. Given the current statistics collected at the Pierre Auger Observatory, the sensitivity of the method is better in Scenario 1 when comparing it to Scenario 2, for the same number of source events above 40\,EeV. Hence, in this work, the method is only applied to events with energies above 40\,EeV.

\subsection{Sensitivity of the thrust method}
\label{sec:sensi-thrust}

In the following the results of the sensitivity studies from the thrust ratio are presented for both introduced scenarios from Section \ref{sec:simu}, with energy thresholds of 20\,EeV and 40\,EeV. Again $1000$ simulations are performed where $N_\text{s}$ events are assigned to the respective source candidate. For each simulation, $100{,}000$ arrival directions were drawn isotropically in every ROI for each of the $1000$ simulations, where the identical number of cosmic rays follow the exposure in the ROI. By comparing the thrust ratio of the ROI with the distribution obtained with the isotropic realizations, we derive the isotropic chance probability $p$.

\begin{figure}
\includegraphics[width=0.45\textwidth]{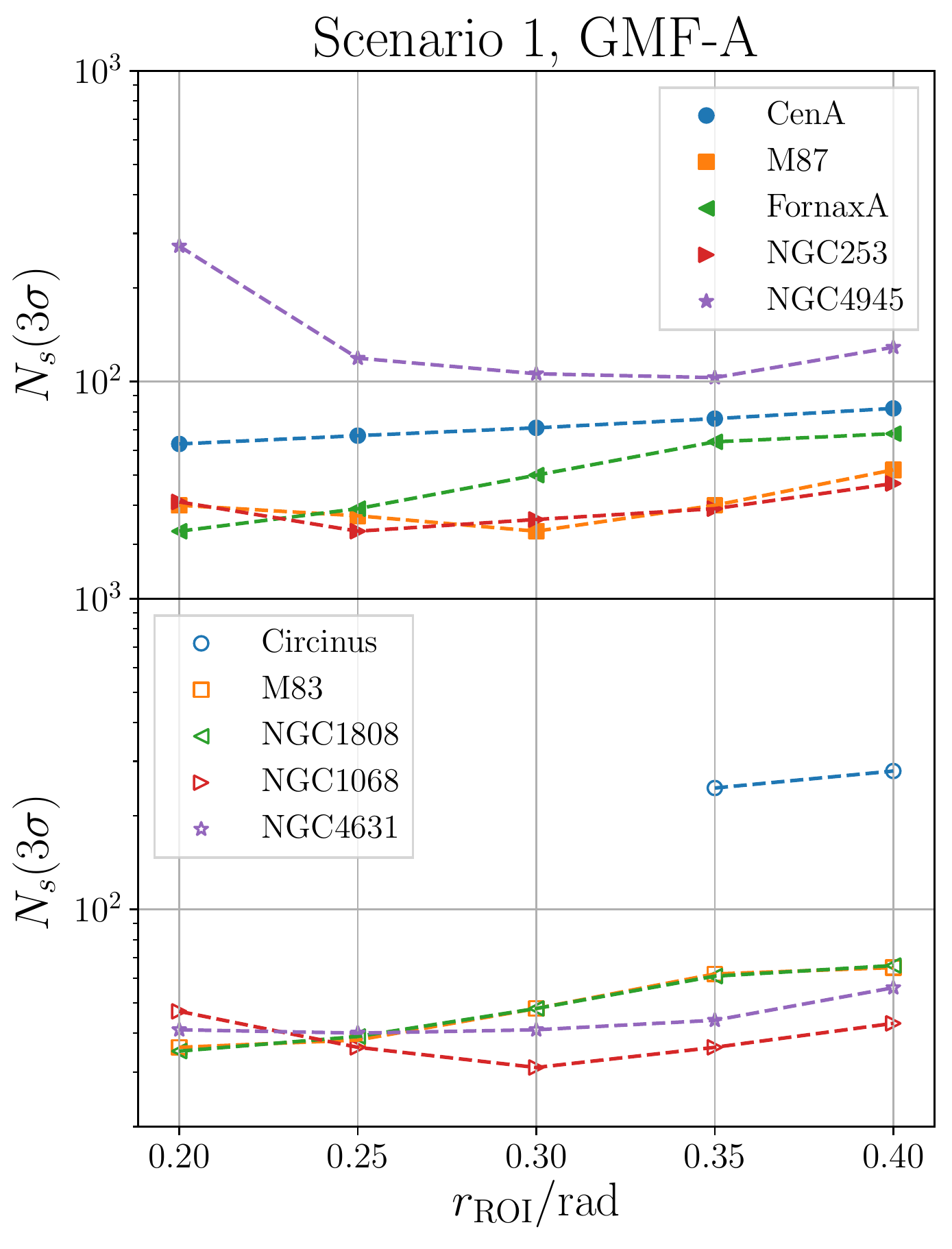}
\hspace{0.5cm}
\includegraphics[width=0.45\textwidth]{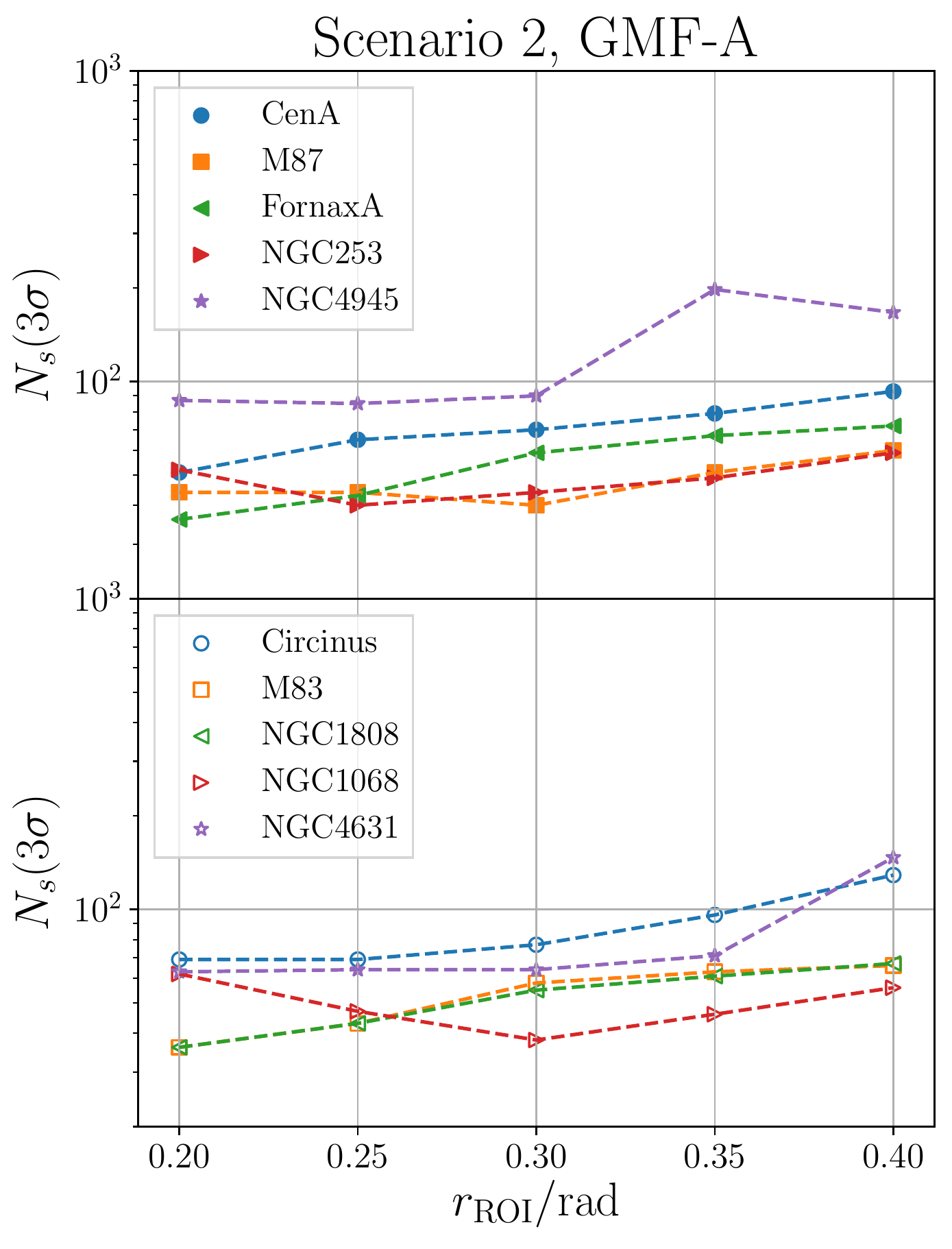}
\caption{Number of injected cosmic ray source events $N_\text{s}$, which leads to an one-sided $3\sigma$ chance probability as a function of the ROI radius $r_{\text{ROI}}$. Applied on the benchmark simulation with the stronger GMF-A model, for Scenario 1 ($E > 40$\,EeV) on the left panel and Scenario 2 ($E > 20$\,EeV) on the right panel.}
\label{fig:hyperparameter_thrust}
\end{figure}

First, we test the sensitivity of the observable with respect to the only free parameter, the radius of the region of interest $r_{\text{ROI}}$. In Figure \ref{fig:hyperparameter_thrust}, we show the number of injected cosmic-ray source events $N_\text{s}$ which leads to a $3\sigma$ significance to reject isotropy as a function of the parameter $r_{\text{ROI}}$ for Scenario 1 in the left panel and Scenario 2 in the right panel. Different colors and markers indicate different source candidates from the catalog in Table \ref{tab:targets}. While there is an overall agreement of the sensitivity curves between Scenarios 1 and 2, the optimal choice for the parameter $r_{\text{ROI}}$ depends strongly on the considered source candidate. This effect is due to the different deflection power of the Galactic magnetic field in the different regions of the sky resulting in patterns that are partially not covered within the ROI for a too small radius $r_{\text{ROI}}$. To obtain an overall good performance, we choose an intermediate radius of the ROI of $r_{\text{ROI}}=0.3$\,rad in the following.

\begin{figure}
\includegraphics[width=0.45\textwidth]{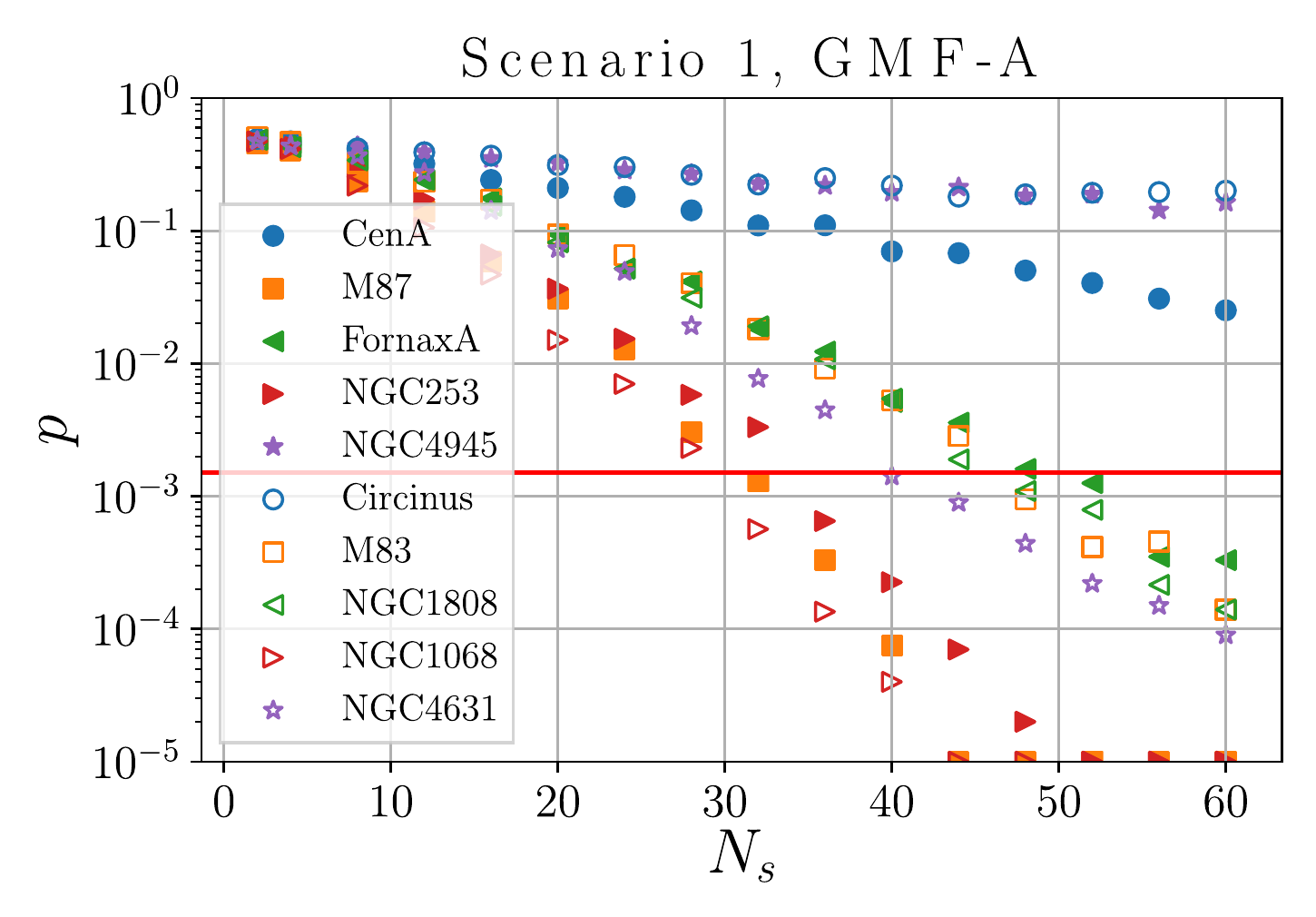}
\hspace{0.2cm}
\includegraphics[width=0.45\textwidth]{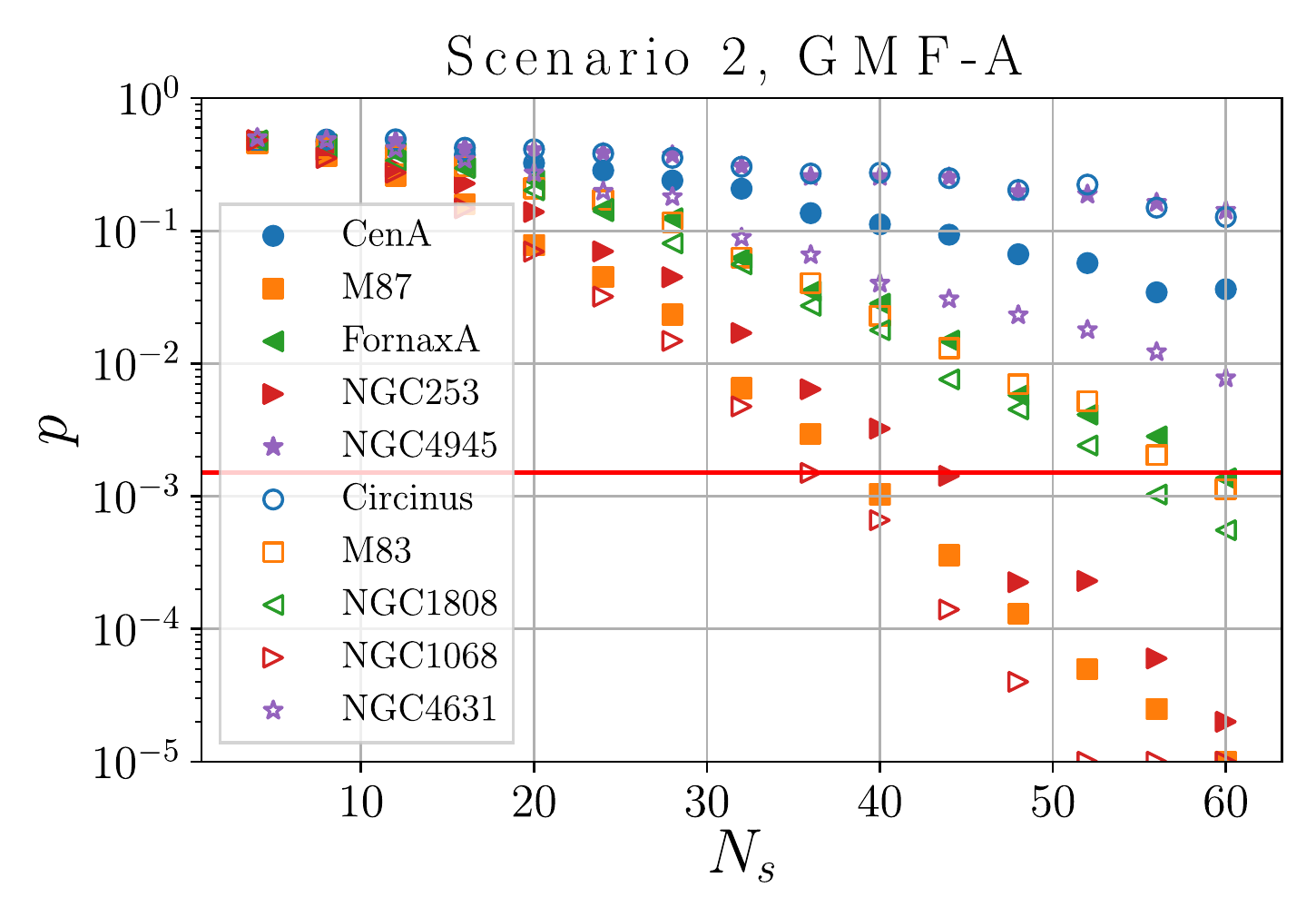}
\caption{Expected sensitivity for the thrust ratio observable on the benchmark simulation with the stronger GMF-A model, for Scenario 1 ($E > 40$~EeV) on the left panel and Scenario 2 ($E > 20$~EeV) on the right panel. The red line indicates the $3\sigma$ confidence level.}
\label{fig:sensitivity_thrust}
\end{figure}

In the next step, we study the sensitivity of the thrust ratio as a function of the number of injected signal cosmic rays $N_\text{s}$. In contrast to the multiplet method, the patterns produced by the source candidates have a larger impact on the sensitivity as can be seen by the large spread between the lines in Figure \ref{fig:sensitivity_thrust}. In case of Scenario 1, with the 40\,EeV energy threshold, a relatively high number of $N_\text{s} \approx 30$ to $N_\text{s} \approx 60$ injected signal cosmic rays is needed to pass the $3\sigma$ confidence level, which corresponds to signal fractions between $3 \%$ and $6 \%$. This performance is inferior to that of the multiplet search, where signal fractions of $1\%$ to $2\%$ are needed to reach the same confidence level. This difference is explained by the fact that the thrust ratio does not make a direct use of the energy information. The weaker turbulent magnetic field model GMF-B just yields a minor improvement on the performance: the same sensitivity as in the case of GMF-A is reached with two fewer signal cosmic rays, on average.

For Scenario 2, with the 20\,EeV energy threshold and the GMF-A model, most of the targets reach the $3\sigma$ confidence level between $N_\text{s} \approx 40$ and $N_\text{s}\approx 60$ signal cosmic rays, corresponding to signal fractions between $0.7\%$ and $1.0 \%$, as can be seen in the right panel of Figure \ref{fig:sensitivity_thrust}. The data above this energy threshold will only be analyzed with this method in the following.

\section{Results of the targeted search}
\label{sec:targeted}
In this Section, we present the results of the targeted search with the source candidates listed in Table \ref{tab:targets}, applying the multiplet and thrust methods to the data collected at the Pierre Auger Observatory described in Section \ref{sec:data}.

The results are shown in Table \ref{tab:targeted-search} for each candidate, for the multiplet search with an energy threshold of 40\,EeV, and for the thrust ratio with energy thresholds of 20\,EeV and 40\,EeV. All the probabilities found with respect to an isotropic distribution of arrival directions are above $1 \%$, thus no significant signature of an alignment pattern around the chosen targets could be found. These probabilities are not penalized for the trial of ten different candidates.

\begin{table}[h]
    \centering
    \caption{Isotropic chance probabilities for the targeted search with the multiplet and thrust ratio methods applied to data collected at the Pierre Auger Observatory. These probabilities are not penalized for the use of ten different candidates.}    \label{tab:targeted-search}
    \vspace{0.25cm}
    \begin{tabular}{ |p{2cm}||p{3.2cm}|p{3.4cm}|p{3.4cm}|  }
    \hline
    \multicolumn{4}{|c|}{Isotropic chance probabilities} \\
    \hline
    Target & \multicolumn{1}{|c|}{Multiplets (40\,EeV)} & \multicolumn{1}{|c|}{Thrust ratio (20\,EeV)} & \multicolumn{1}{|c|}{Thrust ratio (40\,EeV)} \\
    \hline
    Cen\,A   &  $1.2 \times 10^{-2}$ & $0.75$ & $0.42$ \\
    M\,87 &  $0.61$  & $0.44$ & $0.85$ \\
    Fornax\,A & $0.96$ & $0.21$ &  $1.9 \times 10^{-2}$ \\

    \hline
    NGC\,253  &  $0.54$  &  $0.98$  &  $0.88$  \\
    NGC\,4945  & $0.25$ &  $2.9 \times 10^{-2}$  &  $3.7 \times 10^{-2}$ \\
    Circinus  & $0.99$ &  $0.82$  &  $0.58$   \\
    M\,83  & $0.20$ & $0.14$ & $0.54$ \\
    NGC\,4631  & --- &  $0.59$  &  $0.85$  \\
    NGC\,1808  & $0.61$ & $0.63$  &  $0.77$  \\
    NGC\,1068   & $0.75$ &  $6.0 \times 10^{-2}$ &  $0.29$  \\
    \hline
    \end{tabular}
\end{table}

\begin{figure}
\begin{tabular}{cc}
\includegraphics[width=0.45\textwidth]{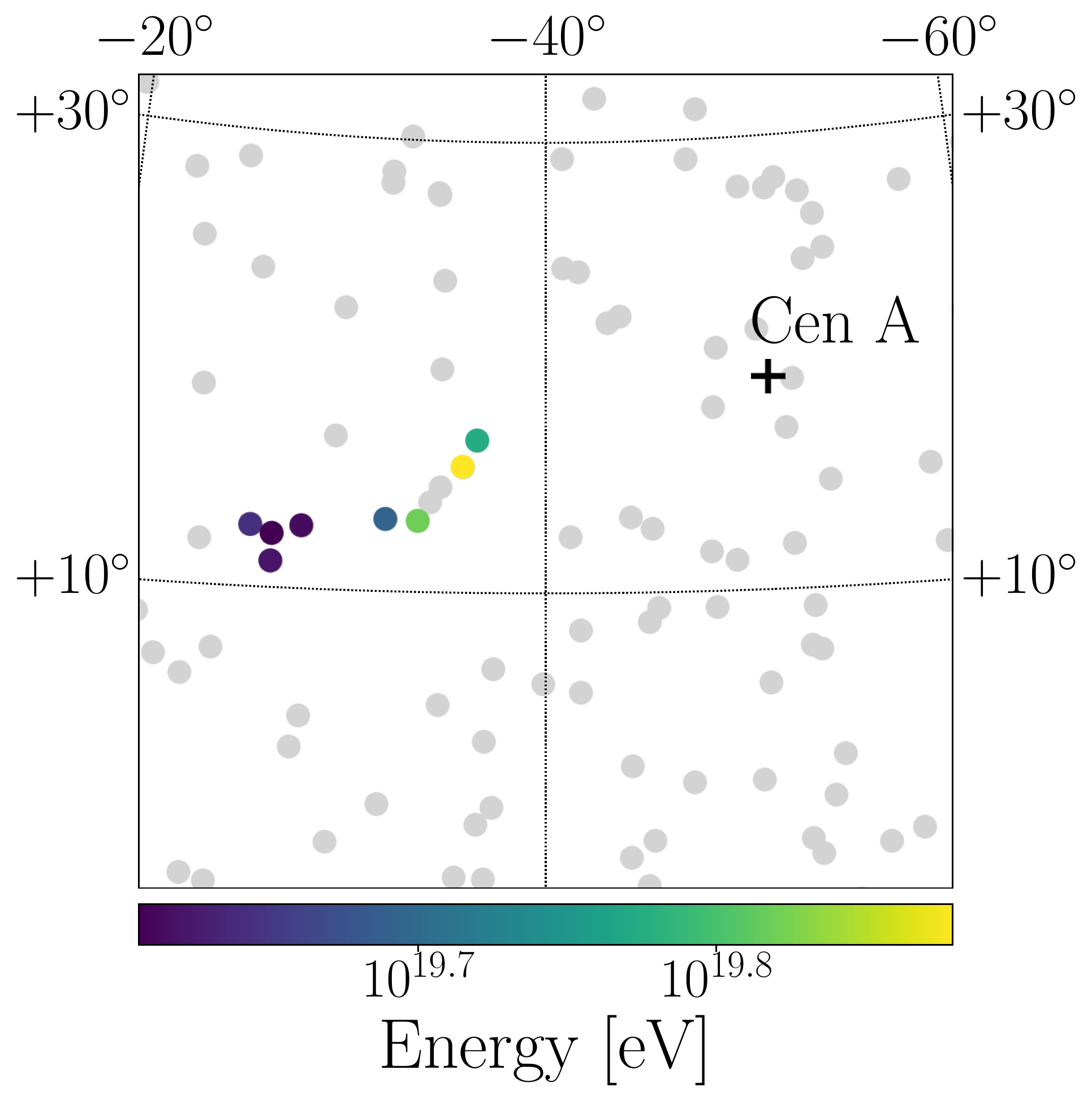} &
\includegraphics[width=0.45\textwidth]{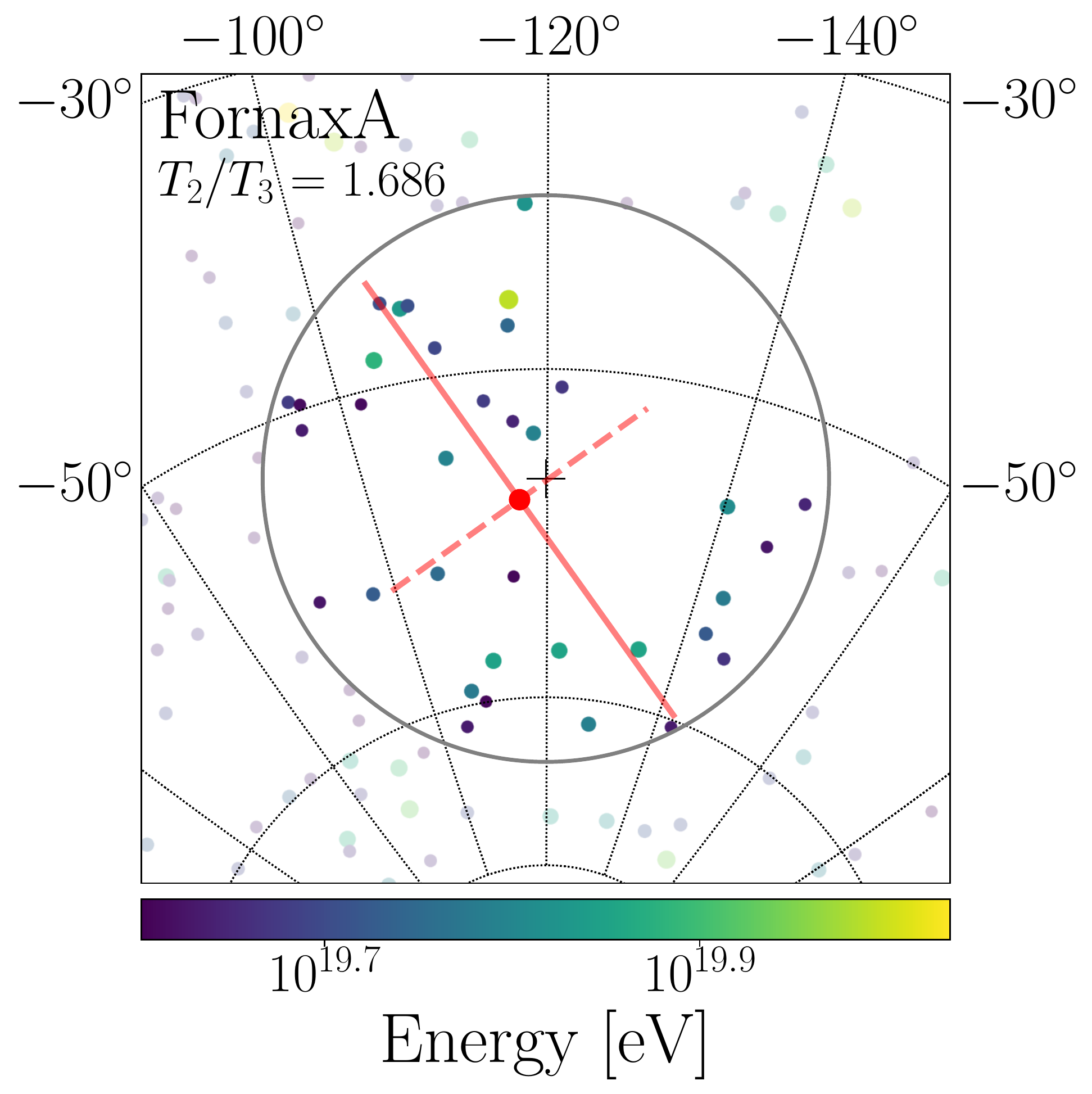} \\
\includegraphics[width=0.45\textwidth]{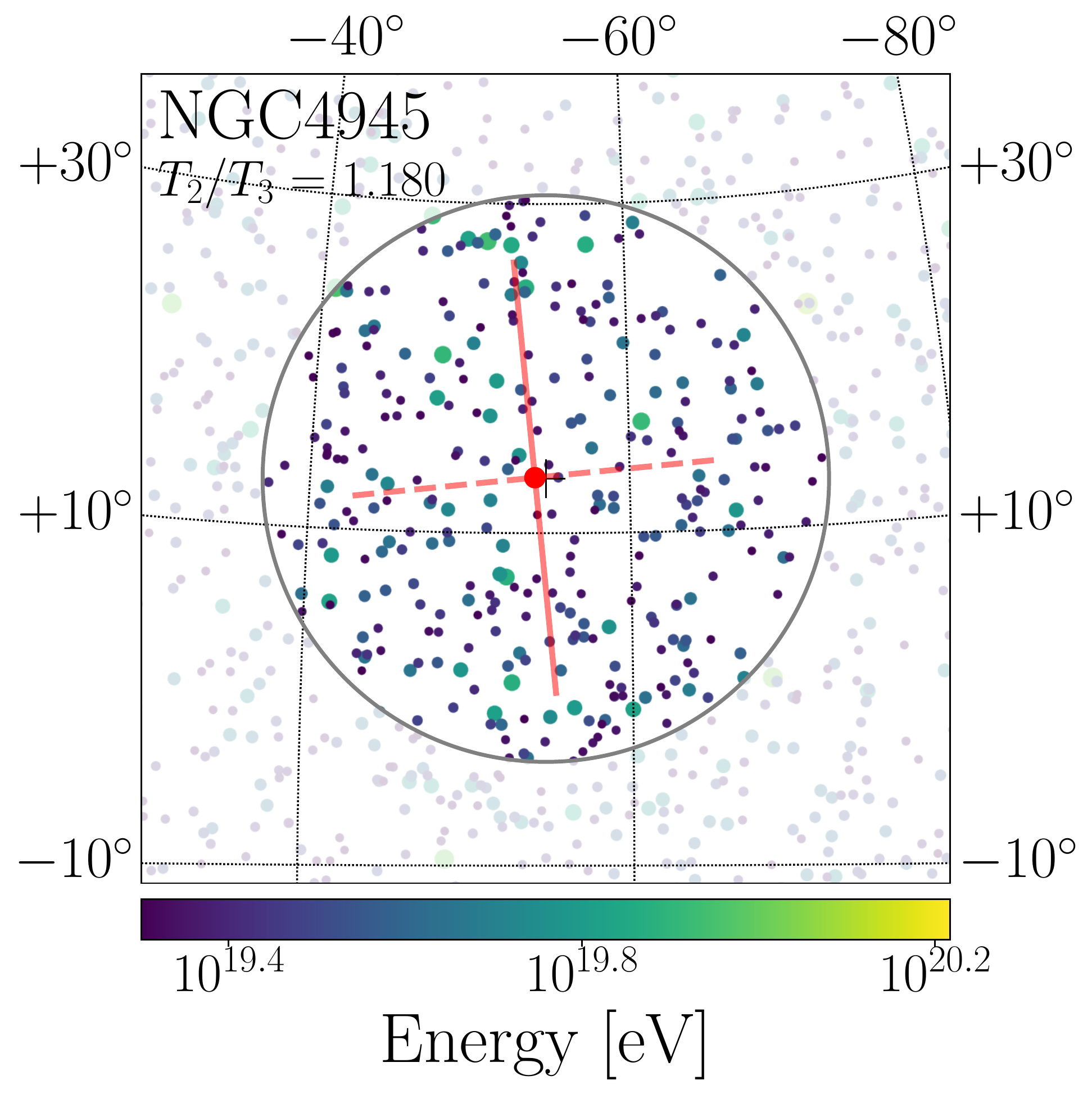} &
\includegraphics[width=0.45\textwidth]{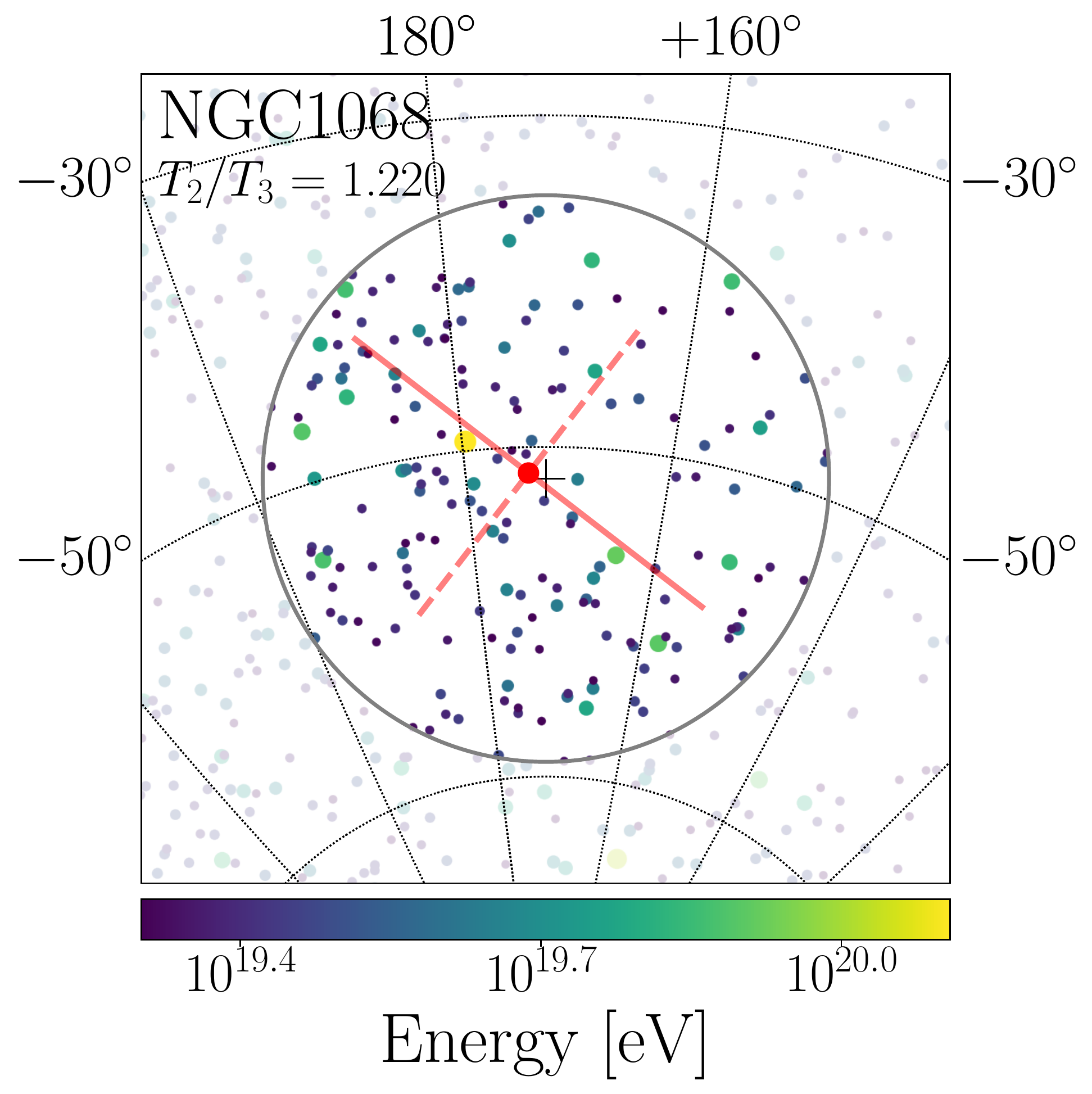}
 \end{tabular}
\caption{Arrival directions of events in Galactic coordinates from chosen sky regions which yield the smallest isotropic chance probabilities in Table \ref{tab:targeted-search}. The upper row shows the AGN candidates Cen\,A with the 8-plet from the multiplet method (left) and Fornax\,A with the thrust ratio (right), both for the energy threshold of 40\,EeV. The lower row shows the SBG candidates NGC\,4945 (left) and NGC\,1068 (right) evaluated with the thrust ratio on the energy threshold of 20\,EeV. The red bars show the thrust vectors $T_2 \cdot \hat{n}_2$ (solid) and $T_3 \cdot \hat{n}_3$ (dashed) as defined in Section \ref{sec:thrust}.}
\label{fig:resultstarget}
\end{figure}

The most interesting but non-significant result for the multiplet search, with $E>40$\,EeV, is with Cen\,A as source candidate, where an 8-plet is found with a chance probability of $1.2 \%$ with respect to isotropic expectations. The fitted deflection power is $D=(9 \pm 2)^{\circ} \times 100$\,EeV. In the upper left panel of Figure \ref{fig:resultstarget}, we visualize the arrival directions of the events that form the 8-plet.

In the case of the thrust ratio observable, the smallest $p$-values are found with the source candidates Fornax\,A (with $E>40$\,EeV), NGC\,4945 (with both energy thresholds) and NGC\,1068 (with $E>20$\,EeV), with chance probabilities between $1.9 \%$ and $6.0 \%$. There is barely any overlap between the cosmic rays within the multiplet of the Cen\,A region and the region of interest evaluated for the thrust observable. In the upper right panel of Figure \ref{fig:resultstarget}, we present the arrival directions of the events in the ROI centered on Fornax\,A above 40\,EeV. The strong response of the thrust ratio ($T_2/T_3 = 1.686$) seems to be an artifact of an asymmetric event distribution in the ROI rather than a promising deflection signature. The lower panel of Figure \ref{fig:resultstarget} shows the candidates NGC\,4945 and NGC\,1068 for energies above 20\,EeV.

It is instructive to look at the lack of magnetically-induced structure in the light of the other measurements carried out at the Observatory. Assuming that the GMF models used in our simulations are a good approximation of the actual one, and considering the anisotropic fractions obtained in \cite{auger2018} for the AGN and SBG models, we should have detected a significant signal in the multiplets and thrust searches if all the events coming from the sources are light elements (p or He). For instance, in the case of Cen\,A, for a total number of 1119 events, ${\sim}76$ events would have been expected from this source. However, the putative multiplet found with Cen\,A as source candidate has a multiplicity of 8 events only. Thus, under these assumptions, the lack of a significant high-multiplicity signal appears to be consistent with a scenario where less than ${\sim}10\%$ of cosmic rays above 40\,EeV are light elements. Such a fraction is in line with our composition constraints at the highest energies \cite{auger2014a,auger2014b}. Overall, the various facets of our data are thus consistent. 

\section{Results of the all-sky blind search}
\label{sec:blind}

We also evaluate both methods on the entire sky, for the multiplet method above energies of $40$\,EeV and for the thrust method for both energy thresholds, $40$\,EeV and $20$\,EeV.

\begin{figure}
\includegraphics[width=0.5\textwidth]{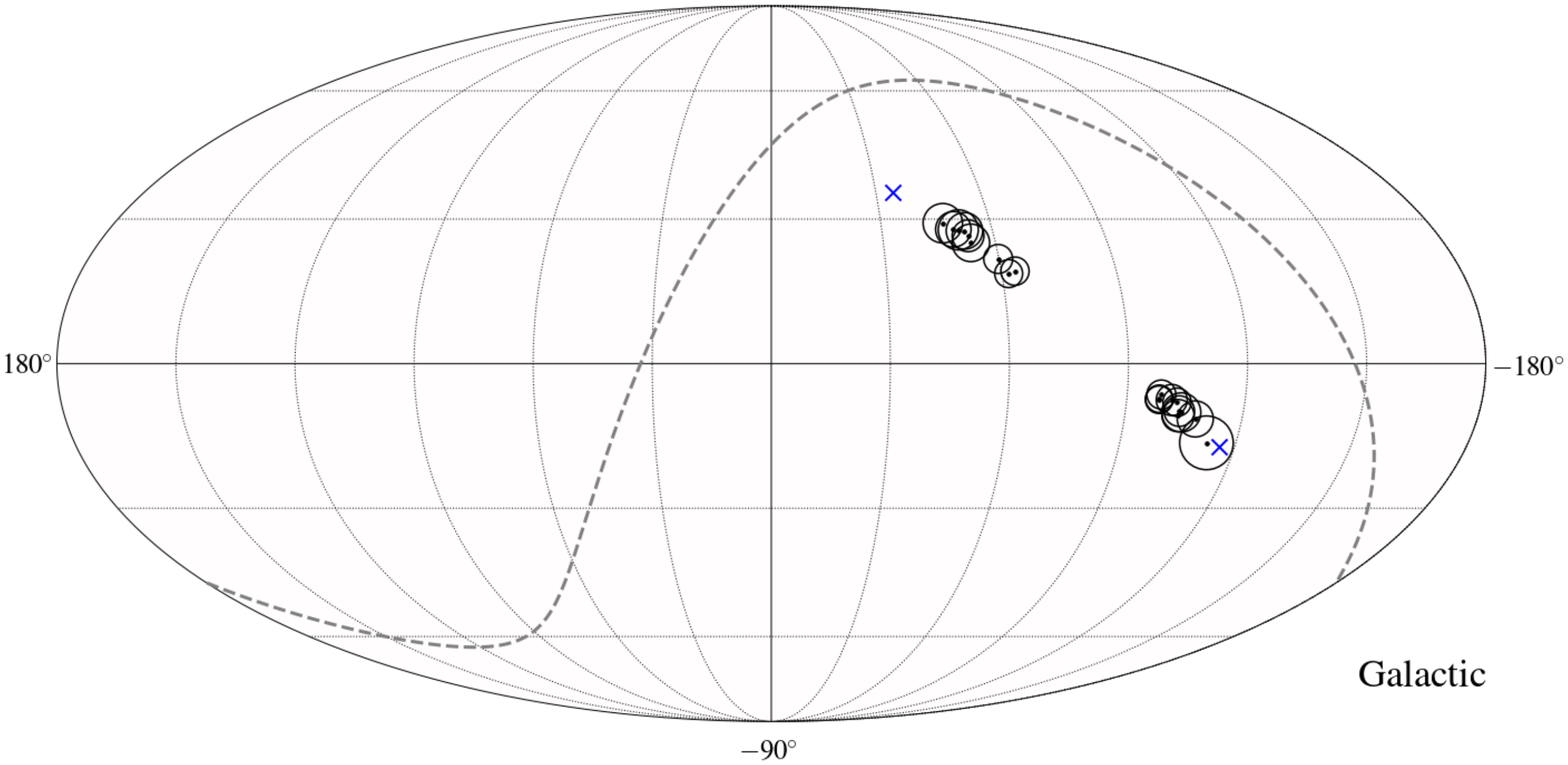}
\includegraphics[width=0.5\textwidth]{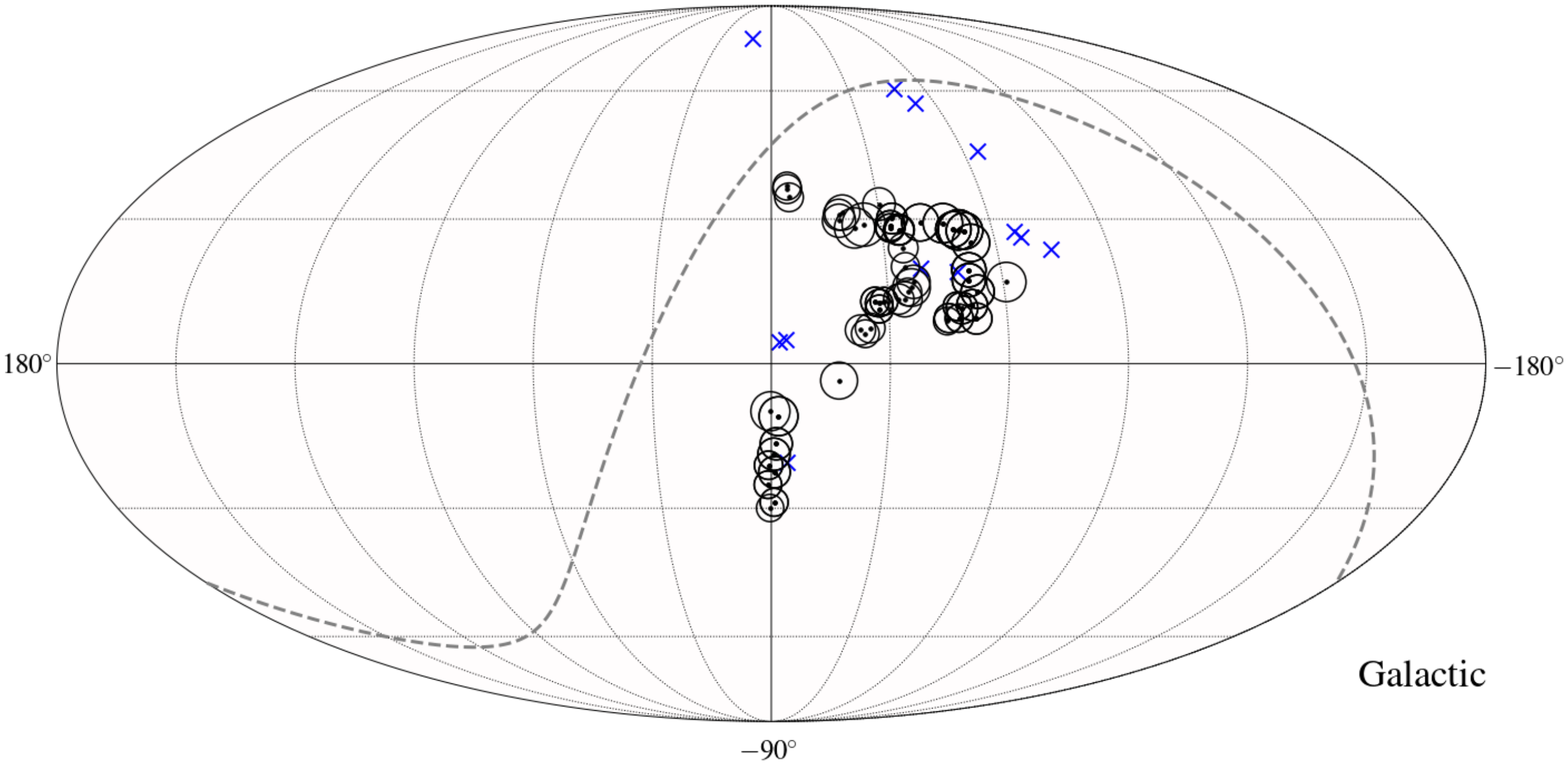}
\caption{Arrival directions in Galactic coordinates of the events that form the 10-plet and 9-plet (left) and 8-plets (right) found on the all-sky blind search with the multiplets method with an energy threshold of $E>40$\,EeV. The radius of the circles is proportional to the energy of the event. Plus signs indicate the positions of the potential sources for each multiplet. The dashed line represents the Equatorial plane.}
\label{fig:resultsblind}
\end{figure}

The largest multiplet found has a multiplicity of $10$ which may appear by chance from an isotropic distribution of events with a probability of $11.4 \%$. The second largest multiplet has a multiplicity of $9$ events. The chance probability of finding at least 2 multiplets with multiplicity larger than or equal to 9 in isotropic simulations is $19.1 \%$. The respective deflection power values are $D=(8.0 \pm 1.3)^{\circ} \times 100$~EeV and $D=(12 \pm 2)^{\circ} \times 100$\,EeV. The arrival directions of the events that form the 10-plet and 9-plet are shown in the left panel of Figure \ref{fig:resultsblind}, as well as the position of what would be their reconstructed source.

When searching for 8-plets, seven independent ones were found. We define ``independent'' multiplets those sets of events that have less than half the number of events in the set in common with others. The chance probability of finding at least 9 multiplets with multiplicity larger than or equal to $8$ in simulations of isotropic arrival directions is $5.6\%$. Thus, no statistically significant results were found. In the right panel of Figure \ref{fig:resultsblind}, we present the arrival directions of the events in all the 8-plets as well as the reconstructed positions of their potential sources.

For the thrust ratio observable, we used the HEALPix scheme \cite{gorski2005} to locate the ROIs equally on the sphere. For each ROI, isotropic realizations with the same number of cosmic rays following the geometrical exposure within the ROI were used to determine local $p$-values. Then, the lowest local $p$-value in the sky is compared with the lowest local $p$-values of $10{,}000$ isotropic skies to determine the post-trial significance. To  overcome computational issues, the isotropic distributions for the thrust ratio observable have been parametrized and pre-calculated on a grid of equatorial declinations and number of cosmic rays within the ROI (cf. Appendix). We select only ROIs with declination lower than $+20^{\circ}$ and a cosmic ray number of at least $15$ to guarantee a sufficiently accurate parametrization.

The lowest but non-significant isotropic chance probability for the thrust ratio observable was found for the energy threshold of 20\,EeV with a ROI centered at Galactic longitude $\ell=-9.8^{\circ}$ and Galactic latitude of $b=-17.0^{\circ}$. The corresponding thrust-ratio was found to be $T_2/T_3 = 1.362$ which translates to a post-trial $p$-value of $25 \%$. For the higher energy threshold of 40\,EeV, we obtained a post-trial $p$-value of $46.7\%$ with a thrust-ratio of $T_2/T_3 = 2.703$. With the ROI centered at $\ell=-151.9^{\circ}$ and $b=-54.3^{\circ}$, we found the same artifact of an unevenly filled ROI close to the Fornax\,A region (cf. Section \ref{sec:targeted}).

\begin{figure}
\includegraphics[width=0.5\textwidth]{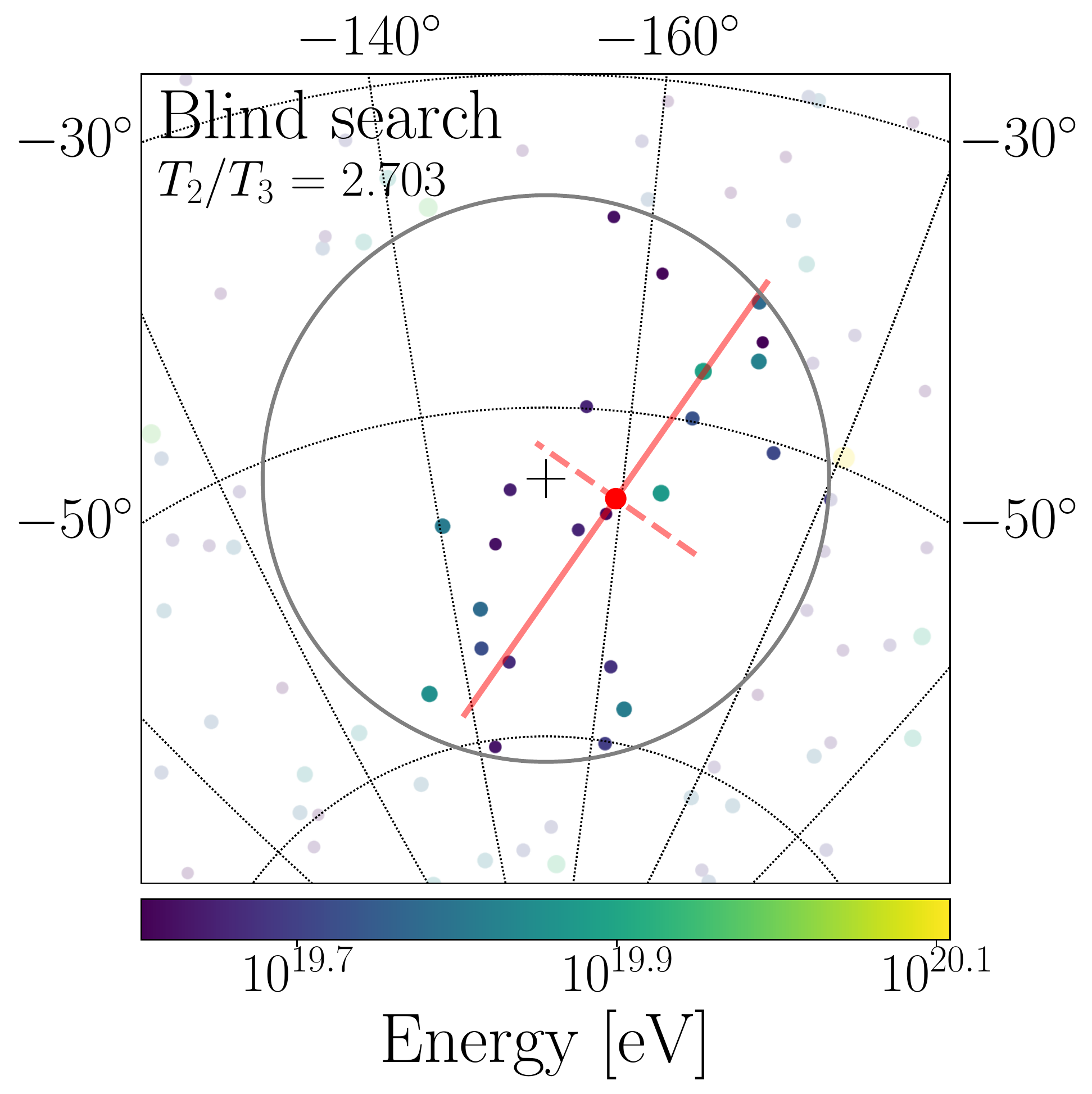}
\includegraphics[width=0.5\textwidth]{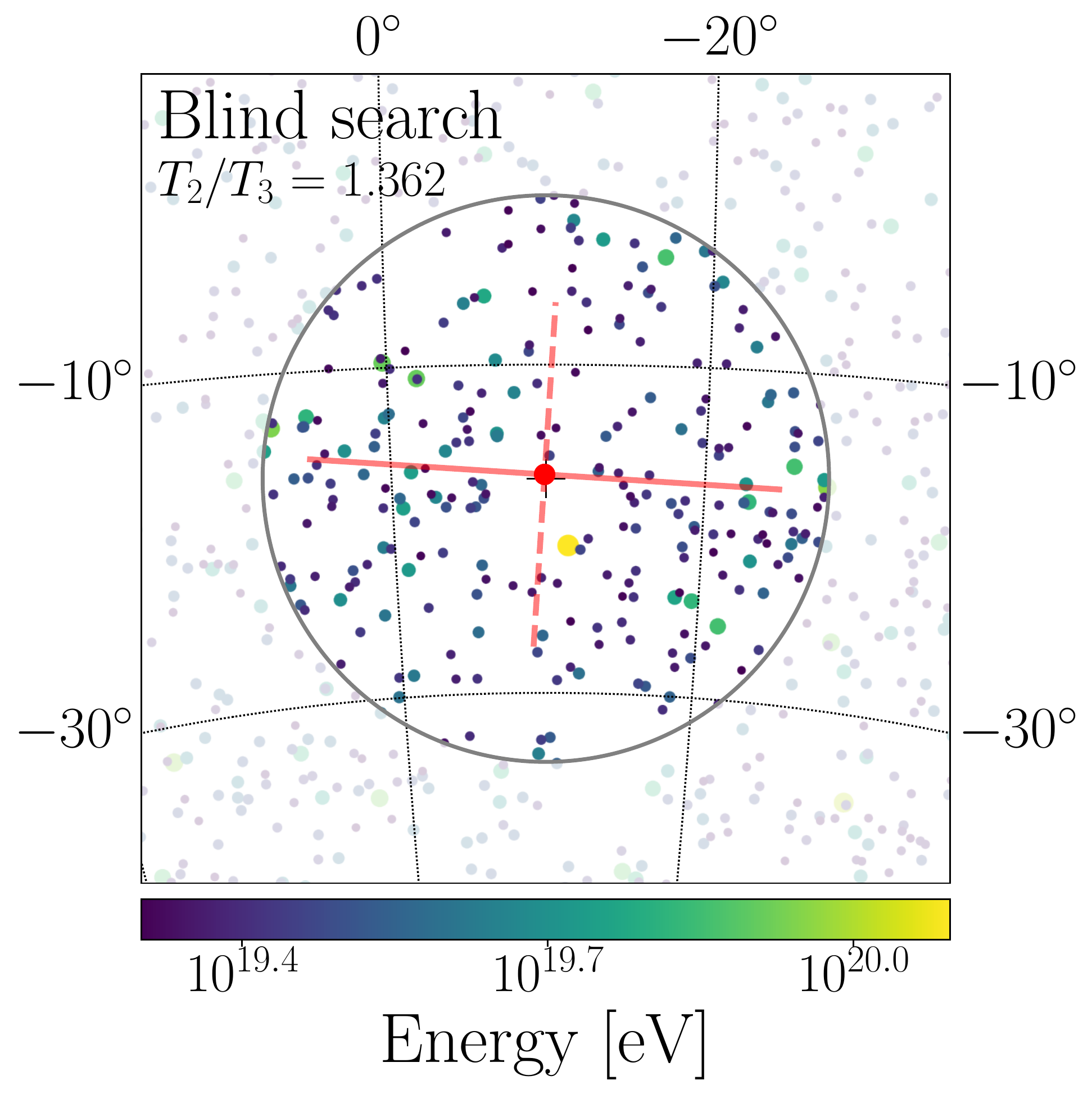}
\caption{Arrival directions of events in Galactic coordinates inside the ROI located in the direction which yields the smallest local $p$-value in the all-sky blind search for the energy thresholds of 40\,EeV (left) and 20\,EeV (right).}
\label{fig:blind_thrust}
\end{figure}

\section{Conclusions}
\label{sec:concl}

Using a set of UHECRs collected at the Pierre Auger Observatory, with an exposure of 101,900\,km$^2$~sr~yr, we searched for indications of magnetically-induced signatures in the arrival directions of the events. We applied two different methods: a search for multiplets and the thrust method. We performed both a search using a list of astrophysical sources, candidates to be sites of UHE acceleration, and an all-sky blind search. We found no statistically significant feature with any of the methods for the energy thresholds considered, 40\,EeV for the multiplets method and 20\,EeV and 40\,EeV for the thrust one. The largest deviations from isotropic expectations were found with a $p$-value of 1.9\% around the Fornax\,A region for the thrust search above 40\,EeV and with a $p$-value of 1.2\% around the Cen\,A region for the multiplet search above 40\,EeV. Given the state of the art of the Galactic magnetic field models, the lack of signal is along the lines of a scenario where less than ${\sim}10\%$ of cosmic rays above 40\,EeV are light elements, which is consistent with our composition constraints at the highest energies.

Future measurements will allow for disentangling of the electromagnetic and muonic components on an event-by-event basis with the upgraded version of the Pierre Auger Observatory \cite{augerupgrade}. Ultimately, the analyses presented in this study will benefit from these measurements which will enable an accurate analysis of mass-sensitive parameters for each event.

\section*{Acknowledgments}

\begin{sloppypar}
The successful installation, commissioning, and operation of the Pierre
Auger Observatory would not have been possible without the strong
commitment and effort from the technical and administrative staff in
Malarg\"ue. We are very grateful to the following agencies and
organizations for financial support:
\end{sloppypar}

\begin{sloppypar}
Argentina -- Comisi\'on Nacional de Energ\'\i{}a At\'omica; Agencia Nacional de
Promoci\'on Cient\'\i{}fica y Tecnol\'ogica (ANPCyT); Consejo Nacional de
Investigaciones Cient\'\i{}ficas y T\'ecnicas (CONICET); Gobierno de la
Provincia de Mendoza; Municipalidad de Malarg\"ue; NDM Holdings and Valle
Las Le\~nas; in gratitude for their continuing cooperation over land
access; Australia -- the Australian Research Council; Brazil -- Conselho
Nacional de Desenvolvimento Cient\'\i{}fico e Tecnol\'ogico (CNPq);
Financiadora de Estudos e Projetos (FINEP); Funda\c{c}\~ao de Amparo \`a
Pesquisa do Estado de Rio de Janeiro (FAPERJ); S\~ao Paulo Research
Foundation (FAPESP) Grants No.~2019/10151-2, No.~2010/07359-6 and
No.~1999/05404-3; Minist\'erio da Ci\^encia, Tecnologia, Inova\c{c}\~oes e
Comunica\c{c}\~oes (MCTIC); Czech Republic -- Grant No.~MSMT CR LTT18004,
LM2015038, LM2018102, CZ.02.1.01/0.0/0.0/16{\textunderscore}013/0001402,
CZ.02.1.01/0.0/0.0/18{\textunderscore}046/0016010 and
CZ.02.1.01/0.0/0.0/17{\textunderscore}049/0008422; France -- Centre de Calcul
IN2P3/CNRS; Centre National de la Recherche Scientifique (CNRS); Conseil
R\'egional Ile-de-France; D\'epartement Physique Nucl\'eaire et Corpusculaire
(PNC-IN2P3/CNRS); D\'epartement Sciences de l'Univers (SDU-INSU/CNRS);
Institut Lagrange de Paris (ILP) Grant No.~LABEX ANR-10-LABX-63 within
the Investissements d'Avenir Programme Grant No.~ANR-11-IDEX-0004-02;
Germany -- Bundesministerium f\"ur Bildung und Forschung (BMBF); Deutsche
Forschungsgemeinschaft (DFG); Finanzministerium Baden-W\"urttemberg;
Helmholtz Alliance for Astroparticle Physics (HAP);
Helmholtz-Gemeinschaft Deutscher Forschungszentren (HGF); Ministerium
f\"ur Innovation, Wissenschaft und Forschung des Landes
Nordrhein-Westfalen; Ministerium f\"ur Wissenschaft, Forschung und Kunst
des Landes Baden-W\"urttemberg; Italy -- Istituto Nazionale di Fisica
Nucleare (INFN); Istituto Nazionale di Astrofisica (INAF); Ministero
dell'Istruzione, dell'Universit\'a e della Ricerca (MIUR); CETEMPS Center
of Excellence; Ministero degli Affari Esteri (MAE); M\'exico -- Consejo
Nacional de Ciencia y Tecnolog\'\i{}a (CONACYT) No.~167733; Universidad
Nacional Aut\'onoma de M\'exico (UNAM); PAPIIT DGAPA-UNAM; The Netherlands
-- Ministry of Education, Culture and Science; Netherlands Organisation
for Scientific Research (NWO); Dutch national e-infrastructure with the
support of SURF Cooperative; Poland -Ministry of Science and Higher
Education, grant No.~DIR/WK/2018/11; National Science Centre, Grants
No.~2013/08/M/ST9/00322, No.~2016/23/B/ST9/01635 and No.~HARMONIA
5--2013/10/M/ST9/00062, UMO-2016/22/M/ST9/00198; Portugal -- Portuguese
national funds and FEDER funds within Programa Operacional Factores de
Competitividade through Funda\c{c}\~ao para a Ci\^encia e a Tecnologia
(COMPETE); Romania -- Romanian Ministry of Education and Research, the
Program Nucleu within MCI (PN19150201/16N/2019 and PN19060102) and
project PN-III-P1-1.2-PCCDI-2017-0839/19PCCDI/2018 within PNCDI III;
Slovenia -- Slovenian Research Agency, grants P1-0031, P1-0385, I0-0033,
N1-0111; Spain -- Ministerio de Econom\'\i{}a, Industria y Competitividad
(FPA2017-85114-P and FPA2017-85197-P), Xunta de Galicia (ED431C
2017/07), Junta de Andaluc\'\i{}a (SOMM17/6104/UGR), Feder Funds, RENATA Red
Nacional Tem\'atica de Astropart\'\i{}culas (FPA2015-68783-REDT) and Mar\'\i{}a de
Maeztu Unit of Excellence (MDM-2016-0692); USA -- Department of Energy,
Contracts No.~DE-AC02-07CH11359, No.~DE-FR02-04ER41300,
No.~DE-FG02-99ER41107 and No.~DE-SC0011689; National Science Foundation,
Grant No.~0450696; The Grainger Foundation; Marie Curie-IRSES/EPLANET;
European Particle Physics Latin American Network; and UNESCO.
\end{sloppypar}

\section*{Appendix}
The distribution of the thrust-ratio observable for an isotropic arrival distribution and an energy spectrum as measured by the Pierre Auger Observatory can be described by the density function

\begin{equation}
 \label{generalized_maxwell_boltzmann}
 f(t; a, b) = \frac{a^{3/b} \, b}{\Gamma (3/b)} \, (t-1)^2 \, \exp (-a (t - 1)^b) \,
\, ,
\end{equation}

\noindent where $t=T_2/T_3$ is the thrust ratio, $\Gamma(x)$ is the Gamma function and $a$ and $b$ are fit parameters. For both scenarios presented in Section \ref{sec:simu}, the density function (\ref{generalized_maxwell_boltzmann}) has been fitted to $T_2/T_3$-distributions as resulting from isotropically distributed cosmic rays. The parameters $a$ and $b$ of this density function depend only on two variables, the equatorial declination $\delta_{\text{ROI}}$ of the ROI center and the number of cosmic rays within the ROI, $N_{\text{ROI}}$. We evaluated $a$ and $b$ of the density function for $30$ different declinations $\delta_{\text{ROI}}$ ranging from $-90^{\circ}$ to $+45^{\circ}$ and for each of them for $25$ different cosmic-ray numbers $N_{\text{ROI}}$ ranging from $-4$ to $+4$ standard deviations in the Poissonian distribution, which depend on the integrated exposure within the ROI. For sufficiently high cosmic-ray numbers $N_{\text{ROI}}$ (above $15$ events), we find that the parameter $a$ is highly correlated to the parameter $b$ as $a = 3.59 \, \exp(0.831\, b^4) + 2.73.$ For any continuous value of the equatorial declination $\delta_{\text{ROI}}$ and the number of cosmic rays $N_{\text{ROI}}$, we use a 2-dimensional interpolation algorithm to determine the parameter $a$ and calculate the local $p$-value with the analytical integral of Equation (\ref{generalized_maxwell_boltzmann}),

\begin{equation}
 \label{generalized_maxwell_boltzmann_pval}
p_{\text{val}} = \int_{t}^{\infty} f(x; a, b) \, \mathrm{d}x = \frac{\Gamma_i(3/b, \, a (t-1)^b)}{\Gamma(3/b)} \,
\, ,
\end{equation}

\noindent where $\Gamma_i(x, \, y)$ is the incomplete upper Gamma function.

\end{document}